\documentclass[fleqn,usenatbib]{mnras}

\usepackage{newtxtext,newtxmath}

\usepackage[T1]{fontenc}

\DeclareRobustCommand{\VAN}[3]{#2}
\let\VANthebibliography\thebibliography
\def\thebibliography{\DeclareRobustCommand{\VAN}[3]{##3}\VANthebibliography}

\usepackage{graphicx}	
\usepackage{amsmath}	
\usepackage{subfigure} 
\usepackage[switch]{lineno}
\usepackage{tabularx} 
\usepackage{longtable}
\usepackage{lscape}
\usepackage{threeparttable}

\makeatletter
\renewcommand\paragraph{\@startsection{paragraph}{4}{\z@}%
            {-2.5ex\@plus -1ex \@minus -.25ex}%
            {1.25ex \@plus .25ex}%
            {\normalfont\normalsize}}
\makeatother
\title[Evolution of M-L Relationship for Filaments]{On the evolution of the observed Mass-to-Length relationship for star-forming filaments}

\author[Feng et al.]{
Jiancheng Feng,$^{1,}$
$^{2,}$
$^{3}$\thanks{E-mail:fengjc@pmo.ac.cn, rjs22@st-andrews.ac.uk}
, Rowan J. Smith$^{4,3}$
, Alvaro Hacar$^{5}$
, Susan E. Clark$^{6,7}$ 
, Daniel Seifried$^{8}$
\\
$^{1}$Purple Mountain Observatory, Chinese Academy of Sciences, 10 Yuanhua Road, Nanjing 210034, China\\
$^{2}$University of Science and Technology of China, No.96, JinZhai Road Baohe District, Hefei, Anhui, 230026, China\\
$^{3}$Jodrell Bank Centre for Astrophysics, Department of Physics and Astronomy, University of Manchester, Oxford Road, Manchester M13 9PL, UK\\
$^{4}$ School of Physics and Astronomy, University of St Andrews, North Haugh, St Andrews, KY16 9SS\\ 
$^{5}$Department of Astrophysics, University of Vienna, T\"urkenschanzstrasse 17, 1180, Vienna, Austria,\\
$^{6}$Department of Physics, Stanford University, Stanford, CA 94305, USA\\
$^{7}$Kavli Institute for Particle Astrophysics \& Cosmology, P.O. Box 2450, Stanford University, Stanford, CA 94305, USA\\
$^{8}$University of Cologne, I. Physical Institute, Z\"ulpicher Str. 77, 50937 Cologne, Germany
}

\date{Accepted XXX. Received YYY; in original form ZZZ}

\pubyear{xxxx}

\begin{document}
\label{firstpage}
\pagerange{\pageref{firstpage}--\pageref{lastpage}}
\maketitle

\begin{abstract}

The interstellar medium is threaded by a hierarchy of filaments from large scales ($\thicksim$100 pc) to small scales ($\thicksim$0.1pc). The masses and lengths of these nested structures may reveal important constraints for cloud formation and evolution, but it is difficult to investigate from an evolutionary perspective using single observations. In this work, we extract simulated molecular clouds from the “Cloud Factory” galactic-scale ISM suite in combination with 3D Monte Carlo radiative transfer code \textsc{POLARIS} to investigate how filamentary structure evolves over time. We produce synthetic dust continuum observations in three regions with a series of snapshots and use the \textsc{Filfinder} algorithm to identify filaments in the dust derived column density maps. When the synthetic filaments mass and length are plotted on an M-L plot, we see a scaling relation of $L\propto M^{0.45}$ similar to that seen in observations, and find that the filaments are thermally supercritical. Projection effects systematically affect the masses and lengths measured for the filaments, and are particularly severe in crowded regions. In the filament Mass-Length (M-L) diagram we identify three main evolutionary mechanisms: accretion, segmentation, and dispersal. In particular we find that the filaments typically evolve from smaller to larger masses in the observational M-L plane, indicating the dominant role of accretion in filament evolution. Moreover, we find a potential correlation between line mass and filament growth rate. Once filaments are actively star forming they then segment into smaller sections, or are dispersed by internal or external forces.

\end{abstract}




\section{Introduction}

A thorough examination of molecular cloud evolution offers insights into the cycle of the interstellar medium (ISM) and star formation across the galaxy, which is a critical field of study in astrophysics \citep{mckee2007theory, kennicutt2012star, krumholz2014big}. 
However, molecular clouds have life-cycles of order 10$^7$ years \citep{heyer2015molecular}, making it challenging to directly track their evolution through observations. Simulations serve as valuable tools to bridge this research area. With the use of simulations, we are able to track the evolution of molecular clouds and investigate the outcomes under various conditions, to study the underlying physics of large scale mechanisms such as turbulence, magnetic fields, and galactic rotation, which are integral to our understanding of star formation and galactic evolution.

Observations in many different tracers reveal that molecular clouds commonly exhibit an internal filamentary structure. 
These filaments can be identified through extinction maps at optical and infrared wavelengths  \citep{schneider1979catalog,hatchell2005star,myers2009filamentary,jackson2010nessie}, as well as through molecular line observations \citep{loren1989cobwebs,mizuno1995overall,schneider2010dynamic}. Far-infrared and submillimeter dust emission maps have also shown the presence of filaments. In particular, the \textit{Herschel} survey revealed the ubiquity of filaments and their importance in the star formation process with dense star-forming cores being distributed along filaments akin to beads strung along a thread \citep{schneider1979catalog,andre2010filamentary,molinari2010clouds,andre2014filamentary}. This illuminates a possible picture where the dense gas within molecular clouds first assembles into dense filaments and then forms dense star-forming cores through fragmentation, as already proposed by \cite{schneider1979catalog}.

It has long been suggested that molecular clouds have a fractal nature, with a hierarchy of structures from clouds to clumps to cores \citep{scalo1990perception,falgarone1991edges,williams1999structure,heyer2015molecular}. 
Similarly, observations show that filamentary structures span several orders of magnitude in scale (see e.g. the review by \cite{hacar2022initial}), from large-scale giant molecular filaments associated with spiral arms on scales of dozens to hundreds of pc \citep{jackson2010nessie,goodman2014bones,wang2015large,zucker2015skeleton,wang2016census,zhang2019star,ge2022census}, to filaments on scales of 1-10 pc \citep{kainulainen2013high,li2016atlasgal,xiong2019co,schisano2020hi}, and even down to fiber structures on scales of 0.1 pc \citep{hacar2013cores,kainulainen2016gravitational,chung2021trao}.
This continuous distribution over such a vast range of scales suggests that filaments possess a hierarchical nature. 
For example, Orion A, when observed at large scales, appears as an elongated filamentary cloud extending up to $\thicksim$ 90 pc \citep{grossschedl20183d}. At intermediate scales, Orion A can be resolved into numerous parsec-size filaments \citep{nagahama1998spatially,johnstone1998jcmt}, while at higher resolutions, small-scale sub-parsec filaments can be identified \citep{wiseman1994fragmentation,Takahashi_2013,hacar2018alma,suri2019carma}. 

Notably, the choice of scale may result in divergent conclusions regarding filament stability. Larger-scale filaments tend to be super-critical, meaning they have line masses exceeding the critical threshold for gravitational collapse, exhibiting greater line masses. In contrast, smaller-scale filaments are more inclined towards sub-critical states, where their line masses fall below this threshold, thus exhibiting smaller line masses \citep{hacar2013cores,henshaw2014dynamical,chen2019filamentary}. Filaments at various scales effectively sample distinct gas densities within the ISM. The mean gas density in molecular filaments has been observed to decrease with increasing length \citep{hacar2022initial}. Parsec-scale filaments typically exhibit average densities of $\thicksim$ 10$^3$ to 10$^4$ cm$^{-3}$, while sub-parsec filaments, commonly referred to as fibres, display densities of > 10$^5$ cm$^{-3}$.

Many theoretical works have sought to investigate the formation and evolution of filaments at different scales.
Turbulence plays a crucial role in the formation of large-scale filament networks.
The origin of the intricate fibres within filaments is a subject of ongoing debate. One proposed scenario, known as the top-down approach, suggests that these structures may arise from their parental filaments, as indicated by the large-scale coherence and parallel organization of fibres with respect to the main filament, as seen in regions like B213-L1495 \citep{tafalla2015chains,clarke2017filamentary}. However, some simulations support an alternative bottom-up mechanism, in which small subsonic filaments initially form within a turbulent medium and are subsequently gathered through collapse and shear flows (\citealt{smith2016nature}).

\cite{hacar2022initial} integrated observations from the past decade to provide an updated description of filaments at different scales and environments, and identified a scaling relation for filament mass and length $L \propto M^{0.5}$, which is derived from the  1st and 2nd Larson relations \citep{larson1981turbulence}. \cite{hacar2022initial} suggest that the distribution of filaments in the Mass-Length diagram might be controlled by how they evolve in terms of a balance between accretion, fragmentation, collapse and destruction. Various studies show velocity gradients tend to be perpendicular to the main axis of the filament, suggesting ongoing gas accretion onto filaments \citep{schneider2010dynamic,beuther2015filament,dhabal2018morphology,williams2018gravity,shimajiri2019probing,chen2020velocity}. The estimated accretion rates, denoted as $\Dot{m}$, typically range from a few 10 to a few 100 $\rm M_{\odot} \ Myr^{-1} \ pc^{-1}$ \citep{kirk2013filamentary,palmeirim2013herschel,schisano2014identification,bonne2020formation,gong2021physical}. Environmental effects play a role in influencing the observed accretion rates, as higher-mass filaments tend to be embedded in regions with higher background column density, which is expect to obtain a higher accretion rate over 500 $\rm M_{\odot}\  Myr^{-1}\  pc^{-1}$ \citep{heitsch2013gravitational,gomez2014filaments,rivera2016galactic}. Moreover, studies suggest that accretion may not stabilize the filaments against gravitational collapse, it can induce fragmentation leading to the formation of dense cores and possibly star formation \citep{mac1998kinetic,hacar2011dense,hacar2013cores,arzoumanian2013formation,seifried2015impact, clarke2017filamentary,heigl2020accretion}.

Accretion along filamentary structures heightens gas density, thereby influencing instantaneous fragmentation properties by reducing the Jeans length of these structures, especially at filament intersection points or along the filament's entire length. Large-scale filaments are inherently turbulent, with supersonic compression creating local over-densities and thus instigating fragmentation before the decay of turbulent energy \citep{mac1998kinetic,mac1999energy}. Numerical simulations further reveal that ongoing material accretion onto the filament significantly impacts gravity-induced fragmentation, deviating notably from the fragmentation of initially quiescent, pre-existing filaments \citep{inutsuka1992self,clarke2016perturbation}. Due to the turbulent energy cascade, accreted material is expected to be moderately turbulent, with fragment locations determined by turbulent motions in the trans- to mildly supersonic regime where fragmentation is shifted from a gravity-dominated to turbulence-dominated process, highlighting the significance of turbulent motions and moderate density enhancements during filament formation  \citep{seifried2015impact,clarke2017filamentary,heigl2020accretion}. \citealt{chira2018fragmentation} suggest that filament fragmentation must occur before reaching the critical line mass. This is because once the filament line mass exceed the critical line mass, the filament will collapse radially in a free-fall time.

This paper is organized as follows. Section 2 introduces the methodology with our simulation, algorithms, and properties calculation. Section 3 presents the basic result of the simulated gas properties. 
We discuss the evolution scenarios and the hierarchical characteristics of filaments in section 4. Our main conclusions are presented in section 5.

\section{Methods}

\begin{figure}
 \includegraphics[width=0.45\textwidth]{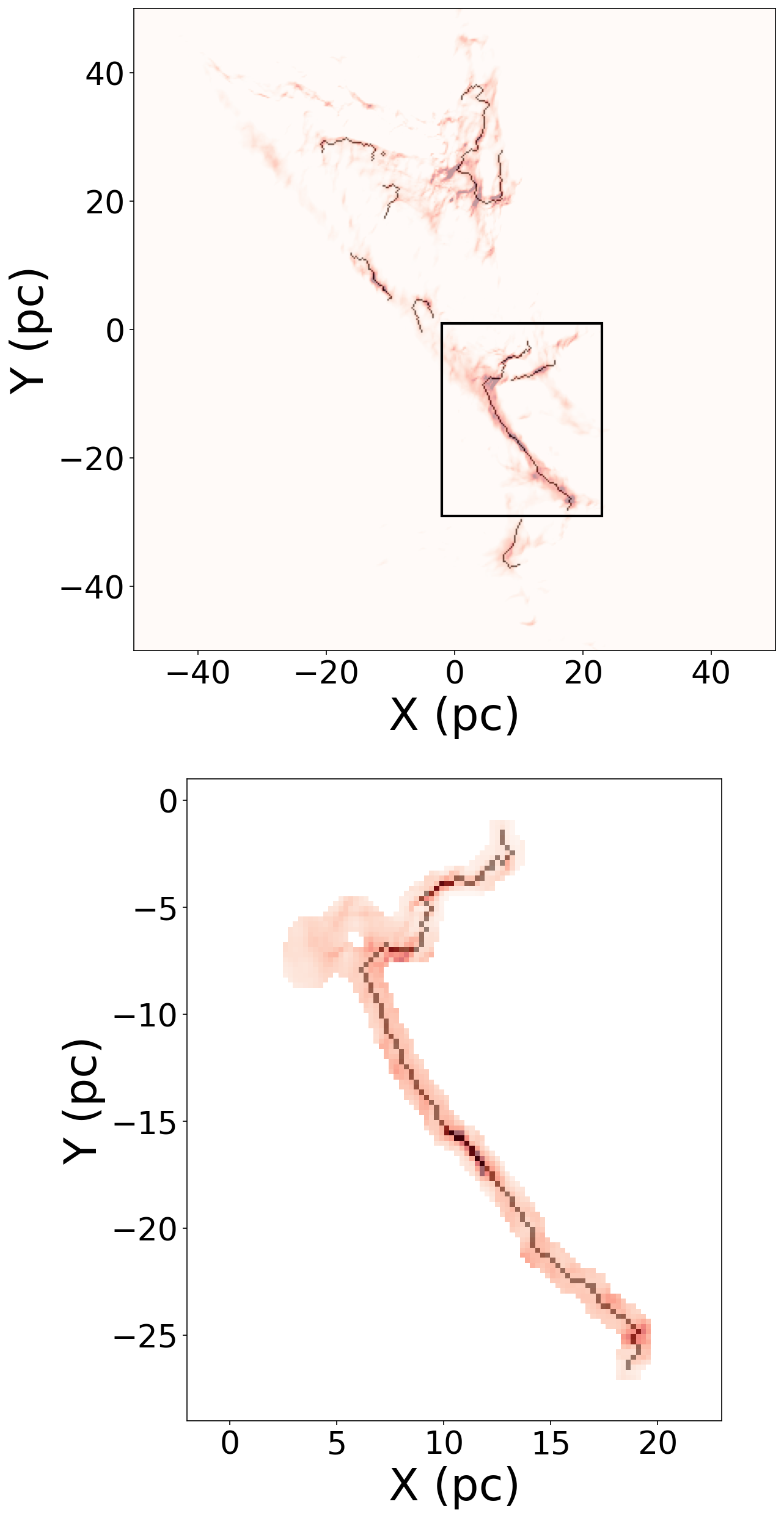}
\caption{An illustration of how \textsc{FilFinder} is used to extract filament RAF2 in snapshot 3. Top: The skeleton of all the identified filaments generated by \textsc{FilFinder} via the column density map. The rectangle shows the area of Filament RAF2 in the column density map. Bottom: The intensity diagram masked by the fitting filament models is used for property calculation. The black line shows the skeleton of the filament. The background shows the $\rm H_2$ column density. The identified filament is first identified in the column density map, and their properties are then derived from the mask and the skeleton.}
 \label{fig:filfinder}
\end{figure}

\subsection{The Cloud Factory simulations}

The Cloud Factory simulations are a suite of high-resolution numerical simulations designed to study the behavior of the interstellar medium (ISM) in a typical spiral galaxy across various scales, ranging from the entire galaxy down to individual filaments and clumps within specific molecular clouds \citep{smith2020cloud}. The Cloud Factory simulation suite utilizes a modified version of the \textsc{AREPO} code \citep{springel2010pur,pakmor2016improving}, which includes various physical and chemical modules that account for different processes occurring in the cold molecular ISM. The model includes a galactic gravitational potential, time-dependent evolution of CO and hydrogen chemistry, ultraviolet extinction, dust absorption, star formation via sink particles, and feedback from supernova explosions.

For the galactic gravitational potential, to minimise computational effort, an analytical approach is adopted to model the large-scale galactic potential. The self-gravity of the gas is calculated using the standard \textsc{AREPO} gravitational tree \citep{springel2010pur}. The axisymmetric component of our analytical gravitational potential is based on the best-fitting potential described in \cite{mcmillan2016mass}, which was developed to align with various observational and theoretical constraints for the Milky Way. We incorporated a spiral perturbation to the axisymmetric potential, following the method outlined in \cite{smith2014nature}, incorporating a four-armed spiral component from \cite{cox2002analytical}.

The chemical evolution of the gas in this study follows the model described in \cite{smith2014nature} and incorporates the hydrogen chemistry framework proposed by \cite{glover2007simulatinga, glover2007simulatingb}. The formation and destruction of CO are treated in a simplified manner based on \cite{nelson1997dynamics}. The hydrogen chemistry model encompasses processes such as H$_2$ formation on grains, H$_2$ destruction via photodissociation and collisional dissociation of atomic hydrogen, H$^+$ recombination in both gas and grain phases, and cosmic ray ionization \cite{glover2007simulatinga}. The evolution of the CO abundance is determined by assuming a limited CO formation rate through an initial radiative association step, and the CO destruction rate is primarily due to photodissociation. Full details of the combined network and a comparison of its accuracy and efficiency with other approaches can be found in \cite{glover2012molecular}.

We utilise the framework of sink particles, which are non-gaseous cells that serve as representations of star formation sites. Cells that surpass a critical density, $\rho_c$, are considered potential candidates for conversion into sink particles. However, these cells must successfully pass a series of energy checks first to confirm whether the gas is unambiguously gravitationally bound and has inwardly directed velocities and accelerations. Moreover, prior to the transformation into a sink particle, a cell must be situated at a local gravitational potential minimum and outside the accretion radius of any existing sink particle. For more details on the methodology used for the creation of sink particles, please refer to the work by \cite{tress2020simulations}. The formation of sinks below protostellar densities implies that not all mass is converted into stars. The formation density and accretion radius in the simulation are determined based on specific criteria and physical constraints. Sink particles are created in cells with densities exceeding a critical density, $\rho_c$. These cells also need to undergo energy checks to ensure they are gravitationally bound and have inward velocities. The accretion radius of the sink particles is initially set to correspond to the Jeans length at their creation density. Over time, the accretion radius increases in order to maintain a constant acceleration at the sink surface. This process ensures that the sinks can effectively accrete mass from neighbouring bound cells. The formation density and accretion radius are adjusted based on the target mass resolution, ensuring accurate representation of star-forming regions and preventing artificial fragmentation in the simulation \citep{greif2011simulations}. We use a star formation efficiency of 33\% in keeping with \cite{matzner2000efficiencies} because we place our sinks at densities typical of protostellar cores. The stellar content of the sinks is obtained by multiplying the sink mass by the assumed star formation efficiency.

To estimate the supernova feedback rate, we use two methods: (1) random supernova explosions, and (2) supernovae tied to sinks. In the first approach, we randomly select points from the initial gas density profile chosen for the disc, assuming a rate of one supernova per 300 years. This is typical for Type Ia supernovae in the Milky Way \citep{diehl2006radioactive}, which should be decoupled from the gas distribution. However, purely random feedback may result in unrealistic cloud properties when considering self-gravity, as it fails to destroy large molecular cloud complexes \citep{gatto2015modelling, walch2015silcc} and does not consider the causal link between massive stars and supernovae. Our secondary feedback method aims to account for this issue by incorporating a randomly distributed supernova component, with a frequency of one supernova every 300 years, as proposed by \cite{tsujimoto1995relative}, to simulate Type Ia supernovae. Additionally, supernovae from sink particles are also included in the model, irrespective of the refinement level at which they were formed. We employ the stellar initial mass function (IMF) as outlined in \cite{kroupa2002initial} and determine the number of high-mass stars within the stellar composition of sink particles based on the methodology described by \cite{sormani2017simple}. As the supernovae from each sink explode individually, this naturally results in clusters of supernova explosions. For further details on our supernovae model, see \cite{tress2020simulations} and \cite{smith2020cloud}.

It also includes a four-armed spiral component and a consistent spiral perturbation to the potential. The gas chemistry description adopts a simplified treatment that assumes a direct conversion between the $\rm C^+$ and CO abundances, following the approach of \cite{nelson1997dynamics}. The non-equilibrium hydrogen chemistry involves reactions between molecular ($\rm H_2$), atomic (H), and ionized ($\rm H^+$) hydrogen, electrons, cosmic rays, dust grains, and the UV radiation field. The star formation model uses a hybrid approach based on sink particles, which can represent either individual stellar systems or clusters of stars.

In this study, three star forming molecular complex regions are chosen from the feedback-dominated Cloud Factory model. We identify the filamentary structures and examine them from an evolving viewpoint using time-evolved snapshots of these regions. These regions are selected based on their distinctive characteristics. Region A is a large molecular complex spanning over 100 pc in size. It is in the early stages of evolution, with barely any sinks. Region B is a relatively small molecular cloud with a low column density and low star formation. Region C contains multiple independent molecular clouds, and it is more evolved than the other two regions.

By examining these three regions, our objective is to span a range of scenarios that are likely to occur for molecular clouds of different sizes and evolutionary stages. The regions are tracked for varying time periods based on their ability to maintain consistency as objects in the time evolution.

\subsection{Radiative Transfer}
\label{sec:RT}
To generate synthetic observations, we perform the radiative transfer simulation using the \textsc{Polaris} code \citep{reissl2016radiative}, which is publicly available. To ensure consistency in resolution and physical quantities, we utilised the same Voronoi grid as the \textsc{AREPO} simulations as in \citet{izquierdo2021cloud}.
We use the precalculated optical properties by \cite{draine2013user}, which consist of a mixture of 62.5\% silicate and 37.5\% graphite. However, this model does not account for the formation of ice mantles around cores, which lead to increased brightness within cores \citep{ossenkopf1994dust}. To address this, we increase the maximum radius to $8\times10^{-6}$ m to obtain a value of $\kappa\sim1$ cm$^2$ g$^{-1}$, consistent with the results from \cite{ossenkopf1994dust} (see also in prep
).

We use a fixed dust temperature, $T_{\rm d}$, as an input parameter for our radiative transfer simulations. This is because the preceding Cloud Factory simulation does not include accretion luminosity or photoionizing feedback, resulting in a dust temperature that was relatively low compared to real observations. To address this, we fix $T_{\rm d}$ to a pumping value of 20 K, which is similar to the peak of the temperature distribution in the probability distribution function of actual observations \citep{lin+2016cloud}. However, this is still a simplified assumption for the dust model. In future work, we plan to improve the sink model to further enhance the accuracy of the dust temperature.

For each snapshot of the target regions, we extract Voronoi grid data from the Cloud Factory, using a default box size of 200 pc as input. We then model dust emission at 70, 160, 250, 350, and 500 $\rm \mu$m by following the Herschel PACS instrument \citep{poglitsch2010herschel} and the SPIRE instrument \citep{griffin2010herschel} guidelines. As input parameters to \textsc{POLARIS}, the specified distance and grid sizes can be easily changed, which provides us with a method to model the same region at different distances, resulting in different spatial resolutions. Specifically, we adopt a fixed angular resolution of $\thicksim$ 20", which is comparable to Herschel observations \citep{hennemann2012spine,lin+2016cloud,arzoumanian2019characterizing}. We choose three distinct distances for the observer detector parameter: 2.5 kpc, 5 kpc, and 10 kpc. This allows us to examine the same region at three distinct physical resolutions, simulating the effects of distance.

The output data is a uniform grid with different space resolutions of 800 $\times$ 800 pixels, 400 $\times$ 400 pixels, and 200 $\times$ 200 pixels, corresponding to 0.25 pc/pix, 0.5 pc/pix, and 1 pc/pix, respectively.

\subsection{Calculation of Properties for Synthetic Dust Observations}
\label{sec:calculation}

In order to derive the properties of the clouds in a manner analogous to observations, we need to perform spectral energy distribution (SED) fitting. SED fitting involves comparing the observed flux densities of an object at different wavelengths with model SEDs that assume a single blackbody component in each line of sight. By fitting the observed SED with a model SED, the physical properties of the object, such as temperature and mass, can be determined. Generally, the SED fitting method requires adjusting the observed flux density in different bands to a common angular resolution that matches the longest wavelength band. In our case, since we use the same resolution for all bands during the radiative transfer simulation, we do not need to perform any smoothing prior to SED fitting.

We use a dust opacity law that is mentioned by \cite{hildebrand1983determination}. This law expresses the flux density $S_{\nu}$ at a given observing frequency $\nu$ as
\begin{equation}
\label{equ:Snu}
S_{\nu}=\Omega B_{\nu}(T_\mathrm{d})(1-e^{- \tau_\nu })
\end{equation}
and,
\begin{equation}
\label{equ:Snu NH2}
N_{\mathrm{H_2}}=\frac{\tau_\nu}{\kappa_\nu \mu m_\mathrm{H}}
\end{equation}
$B_{\nu}$ is the Planck function for a given temperature $T_\mathrm{d}$, derived by
\begin{equation}
\label{equ:Bnu NH2}
B_{\nu}(T_\mathrm{d})=\frac{2h\nu^3}{c^2} \frac{1}{exp(h\nu/kT_\mathrm{d})-1}
\end{equation}
where
\begin{equation}
\kappa_\nu = \kappa_{230} (\frac{\nu}{\mathrm{230GHz}})^\beta
\end{equation}
is the dust opacity, $\beta$ is the dust opacity index, $\Omega$ is the considered solid angle, $\mu = 2.8$ is the mean molecular weight, $m_\mathrm{H}$ is the mass of a hydrogen atom, and $\kappa_{230} = 0.09 \ \rm cm^2 \ g^{-1}$ is the dust opacity per unit mass at 230 GHz. We interpolate the opacity $\kappa_{\nu}$ from \cite{ossenkopf1994dust}. A gas-to-dust mass ratio of 100 is assumed.

We use the method described in \cite{lin+2016cloud} to fit the molecular hydrogen column density $N_{\mathrm{H_2}}$ with all the radiative transfer simulation bands (70, 160, 250, 350, and 500 $\rm\mu m$). We fix the value of $T_\mathrm{d}$ equal to 20 K as discussed in Section \ref{sec:RT}, therefore, the $T_\mathrm{d}$ has no physical meaning, and we only consider the $N_{\mathrm{H_2}}$ result for subsequent analysis.


After deriving the maps of $N_{\mathrm{H_2}}$, we define the boundaries of the clouds by creating masks for each region where $N_{\mathrm{H_2}}$ is greater than $10^{21}$ $\rm cm^{-2}$. We then calculate the gas mass for each pixel using the $\rm H_{2}$ column density maps, based on the formula
\begin{equation}
\label{equ:mass calculate}
M=\mu m_{\mathrm{H}}D^{2}\int{N(H_{2}) d\Omega}
\end{equation}
where $\mu$ is the mean molecular weight per hydrogen molecule, which is assumed to be 2.8 in this work. The mass of atomic hydrogen is represented by $m_{\mathrm{H}}$, and D represents the distance to the object. The solid angle element is denoted as $d\Omega$. Finally, we sum all the pixels within the mask to obtain the total 'observed' gas mass of the molecular clouds. Additionally, we calculate the mean $N_{\mathrm{H_2}}$ within the boundary mask. The calculated gas mass is denoted as $M_{\rm H_{2,dust}}$ for synthetic dust data and $M_{\rm H_{2,raw}}$ for original projection data, while $M_{\rm H_{2,raw}}$ is calculated with the projection $\rm H_2$ column density map from the original simulation data.
In the Cloud Factory simulations, dense star-forming cores are represented by sink particles. These particles are associated with both a stellar mass $M_{\rm stellar}$ (33\% of the sink mass) and a gaseous mass $M_{\rm sinkgas}$ (67\% of the sink mass). Consequently, the H$_2$ mass derived from radiative transfer calculations misses a significant portion of the total cloud mass. To address this issue, when calculating the total mass for the filament, we consider the'missing' gaseous component of the corresponding sink mass, in addition to the molecular gas mass derived from radiative transfer column density maps. Additionally, we examine the discrepancies between the raw datasets and the data produced by \textsc{POLARIS}. The detailed results can be found in Appendix \ref{sec:appendix}. Notably, there is a good alignment in column density between the two datasets within the dense gas regions.

\subsection{Filaments identification and Profile Calculation}
\label{sec:filfinder}


 \begin{table}
 \caption{$\rm H_2$ column density threshold for \textsc{FilFinder} filament identification and the corresponding snapshot series of three regions. }
  \label{tab:filfinder}
    \resizebox{\linewidth}{!}{
  \begin{tabular}{ccc}
    \hline
    Region & N$\rm _{H_{2}}$  Threshold & snapshots$^a$ \\
         &     10$^{21}$ cm$^2$  &  \\
    \hline    
    A  &   1.7    & 3,8,13,18,23   \\
    B  &   0.5    & 3,4,5,6   \\
    C  &   4.5    & 3,5,7,9,11,13,15   \\
    \hline \\
    \multicolumn{3}{l}{$^a$ The number of the snapshots series indicate the evolving time with a unit of $10^5$ yrs. }\\
  \end{tabular}}
 \end{table}

In this work, we primarily use the \textsc{FilFinder} algorithm \citep{koch2015filament} to identify structures.
This extracts filaments from a column density map by creating a mask beyond an adaptive column density threshold, derived from the intensity image, and then reducing the mask to a skeleton via the Medial Axis Transform. It returns all the possible filamentary structures across the input map and is able to extract hierarchical filament structures uniformly \citep{zhang2021algorithm,zucker2018physical,liu2018liu,xiong2019co}. However, \textsc{FilFinder} can only be applied to 2D diagrams, although \cite{kim2023kinematic} recently presented a method for identifying filaments in 3D position-position-velocity data based on applying FilFinder to velocity channel maps.

We apply \textsc{Filfinder} on the H$_2$ column density maps originating from the RT simulated dust emission, as described in Section \ref{sec:calculation}. We first define a setting of 'flatten\_percent=99.9' to emphasise the main structure. This means that during image contrast adjustment, intensities up to the 99.9th percentile of the data are retained. This approach helps to minimise the impact of extreme brightness values in the image, which can interfere with subsequent image processing steps like filtering and feature detection. As the clouds are in different evolutionary states, their density varies significantly, so rather than choosing a single threshold, we aim to create a mask with a threshold over the top 1\% of the data for each region such that only the dense structure is included.
This is similar to real observations, where no universal threshold defines a filament. Instead, different thresholds are applied to various regions based on the column density contrast with their environment, aiming to highlight the morphological features.

The value of the threshold for each mask is shown in Table \ref{tab:filfinder}. The column density threshold is kept constant for each region for different snapshots. It should be noted that this threshold is relatively low and is only used for a pre-process to maintain most of the structures. Using the mask, the algorithm generates skeletons and branches. The shortest lengths of these are set by two variables: "skel$\_$thresh" and "branch$\_$thresh". "skel$\_$thresh" is set to 8 pc, which corresponds to 8, 16, and 32 pixels in three respective resolutions. On the other hand, "branch$\_$thresh" is set to 4 pc, corresponding to 4, 8, and 16 pixels in the three resolutions, respectively. The skeleton and the branches trace the bones of the filament. The skeleton is the longest part of the main structure, and the branches are shorter parts connected to the skeleton.
Here, we only focus on the skeleton part of the filaments when defining the length. The algorithm also derives the widths of the filaments, which are fitted by a Gaussian with a mean fixed to 0 and a constant background.

With the length and width determined, the algorithm also creates a boundary mask of filaments containing both skeleton and branches with a threshold larger than $10^{21}\rm \  cm^{-2}$, as shown in Figure \ref{fig:filfinder}. This mask enables the calculation of the total mass of the filaments from the input image. The length of the filament is computed as the product of the total pixel number of the identified skeleton and the physical length of a pixel. Using the modelled skeleton and width, we generate the filament mask and calculate the mass of each pixel from the column density map. Subsequently, we sum up the mass of the pixels within the mask to obtain the mass of the filament. Finally, we extract the sink data from the original Cloud Factory dataset, where each sink comprises 67\% of the gas portion and 33\% of the stellar mass. If the sink position is within the filament mask, we then sum up the mass derived from the radiative transfer dust image and 67\% of the sink mass to obtain the total mass of each filament.

\section{Results}

\begin{figure*}
\includegraphics[width=\textwidth]{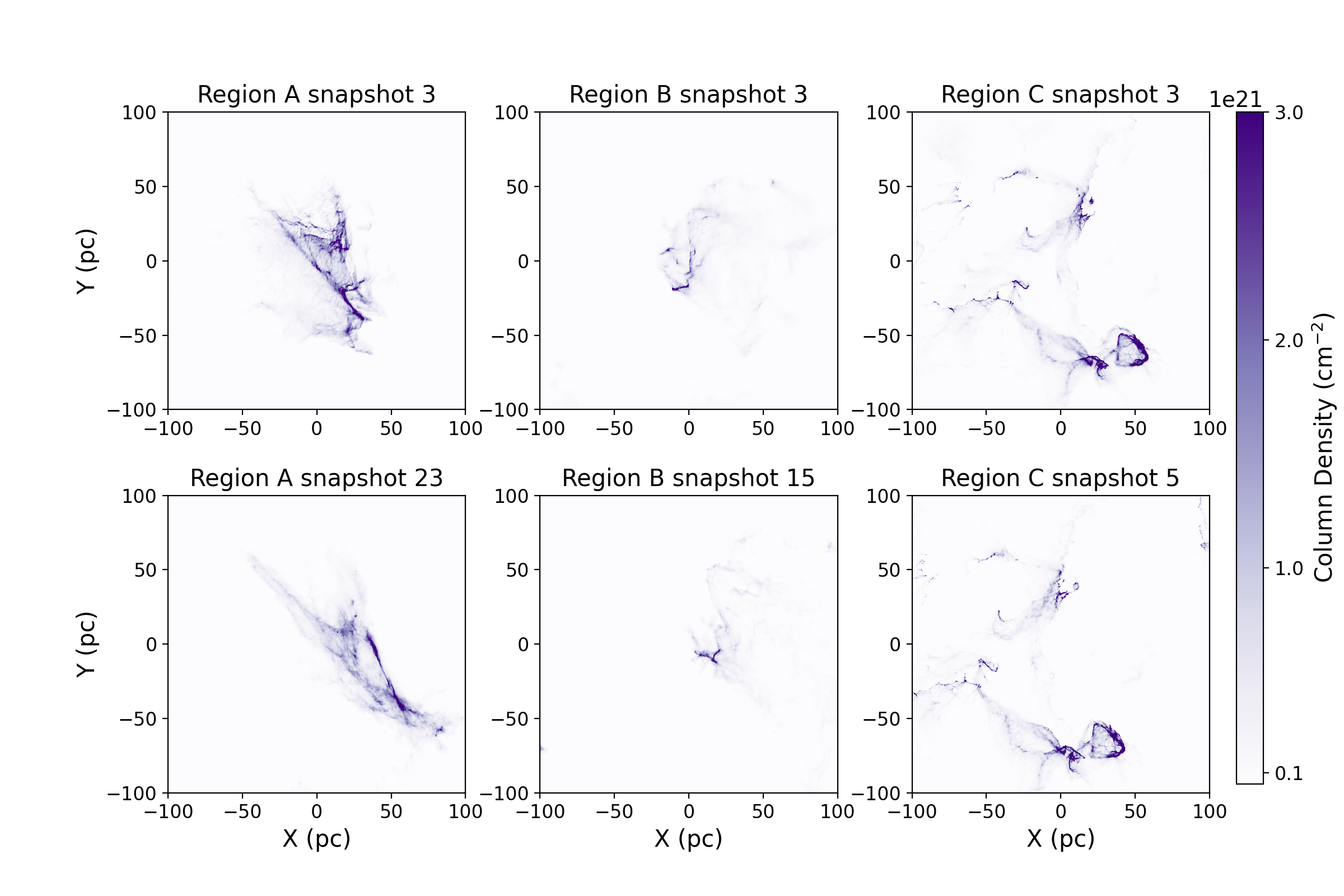}

\caption{$\rm H_2$ column density maps derived from radiative transfer dust image with 0.25 pix/pc resolution. The top panels are the first snapshot of three regions, and the bottom panels are the last snapshot of three regions. The background shows the $\rm H_2$ column density. The column density maps show the morphology change between the initial snapshots and the final snapshots.} 
\label{fig:cloud nh2 maps}
\end{figure*}

\begin{table}
\caption{The gas properties in three regions, as derived from all snapshots of the highest resolution (0.25 pc/pix) synthetic observational data and the original simulation data.}
\begin{threeparttable}

{\small
\setlength\tabcolsep{6pt} 
\begin{tabular}{cccccc}
\hline
Region & Snapshot & $N_{\rm H_{2}}$ & $M_{\rm H_{2,dust}}$  & $M\rm _{Sinkgas}$ & $M\rm _{Stellar}$  \\
 \hline
 &  & $10^{20}$ cm$^2$ & M$_{\odot}$ & M$_{\odot}$ & M$_{\odot}$ \\
        (1) & (2) & (3) & (4) & (5) & (6) \\ \hline
        A & 3 & 5.91 & 71680 & 96 & 64 \\ 
        A & 8 & 6.44 & 71339  & 1014 & 676 \\ 
        A & 13 & 6.57 & 68008  & 1500 & 1000 \\ 
        A & 18 & 6.28 & 65333  & 1796 & 1197 \\ 
        A & 23 & 5.77 & 63101  & 2028 & 1352 \\ 
        B & 3 & 2.61 & 25337  & 1706 & 1137 \\ 
        B & 5 & 2.61 & 24105  & 2414 & 1609 \\ 
        B & 7 & 2.59 & 22355  & 3620 & 2413 \\ 
        B & 9 & 2.65 & 20440  & 4572 & 3048 \\ 
        B & 11 & 2.65 & 18590  & 5249 & 3499 \\ 
        B & 13 & 2.60 & 17409  & 5868 & 3912 \\ 
        B & 15 & 2.61 & 16445  & 6302 & 4201 \\ 
        C & 3 & 4.52 & 71442  & 18024 & 12016 \\ 
        C & 4 & 4.54 & 68925  & 21027 & 14018 \\ 
        C & 5 & 4.61 & 67708  & 23501 & 15667 \\ 
        C & 6 & 4.80 & 62836  & 28485 & 18990 \\
 \hline
\end{tabular}
}
\begin{tablenotes}
\item[] \textbf{Note.} (1) Region of clouds (2) Evolving time snapshot, in units of 10$^5$ years (3) The mean $\rm H_2$ column density of clouds derived by SED fitting via dust emission (4) Total mass of all clouds in this region, derived from dust emission (5) Total mass of the gas portion in sinks (6) Total mass of the stellar portion in sinks.
\end{tablenotes}
\label{tab:clouds properties}
\end{threeparttable}
\end{table}

\subsection{Cloud Properties}
\label{sec:regions}

Before identifying filaments within our dataset, it is important to determine how the gas around the filaments affects them. Figure \ref{fig:cloud nh2 maps} displays the synthetic $\rm H_2$ column density maps for the first and last snapshots of each of the three regions. Table \ref{tab:clouds properties} summarises the column densities and masses for the synthetic observations with a resolution of 0.25 pc/pix, which were calculated using the methodology described in Section \ref{sec:calculation}. Notably, the molecular gas mass $M_{\rm H_{2,dust}}$ in Table \ref{tab:clouds properties} excludes the portion of gas within the sinks, as the radiative transfer simulation is restricted to modelling gas that has not transitioned into a sink.

We trace different durations and use varying time steps for the snapshots of these three regions, due to their distinct stellar evolutionary stages. For Region A, as indicated in Table \ref{tab:clouds properties}, the stellar mass, $M_\mathrm{{stellar}}$, in the initial snapshot 3 is significantly lower than in the other regions, with almost no star formation. As a result, the evolutionary duration of Region A is the longest, spanning 2 Myrs. In contrast, Region C has a substantial amount of star formation right from its initial snapshot, leading us to track it for the shortest duration of 0.4 Myrs. This is because of the omission in our simulation of early stellar feedback components, including jets, outflows, and photoionization. Consequently, filament evolution becomes distorted after significant star formation, making further evolutionary tracking less meaningful. Region B is intermediate between the two, with a tracked duration of 1.2 Myrs. In this case, we opted for different time steps to ensure a comparable number of snapshots across the three regions.

Region A comprises a complex molecular system with a mass of molecular gas $M\rm_{H_{2,dust}} \approx 2 \times 10^{5} \ M_{\sun}$ in the masked region. We examine 5 snapshots, 3, 8, 13, 18, and 23, representing the time (in increments of 10$^5$ years) since we zoom into the region in the Cloud Factory simulation. However, due to its complex molecular structure, it is difficult to track the evolution of the individual identified filaments in Region A from beginning to end, as these filaments interact and even intertwine with each other.

Region B consists of a single cloud that we analyse from snapshot 3 (0.3 Myrs) to snapshot 15 (1.5 Myrs) with a time step of 0.2 Myrs. The total gas mass in this region is $M\rm_{H_{2,dust}} \approx 3 \times 10^{3}\  M_{\sun}$. The column density of Region B is lower compared to the other regions, and we can track one particular filament over a longer time period because of the region's simplicity.

Region C contains multiple isolated filamentary objects, making it a suitable case for tracing the average properties of multiple filaments evolving over time. The total mass of Region C is $M\rm_{H_{2,dust}} \approx 3 \times 10^{5} \ M_{\sun}$. We analyse four snapshots, from snapshot 3 (0.3 Myrs) to snapshot 6 (0.6 Myrs).

\subsection{Filament properties at different resolutions
}
\label{sec:filament description}

\begin{figure*}
\includegraphics[width=\textwidth]{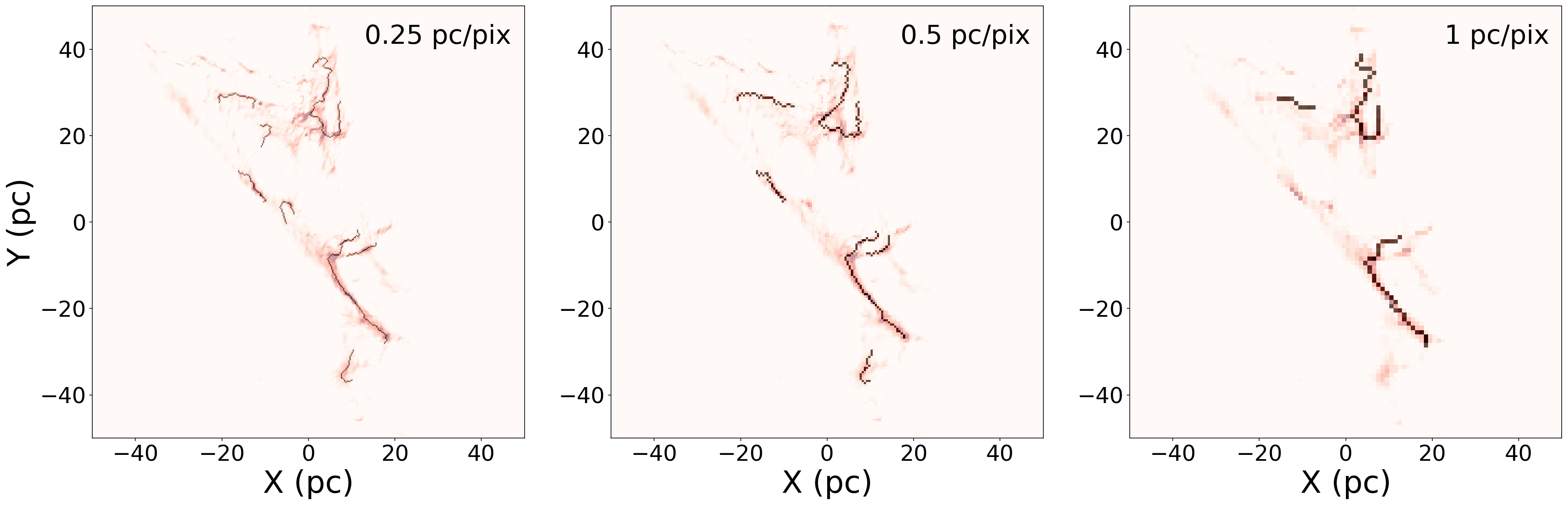} 

\caption{The identified \textsc{FilFinder} skeleton for Region A of snapshot 3 for different resolutions. The background shows the integrated intensity diagram, and the black lines show the identified filament skeletons. The resolution of each panel is labelled in the top-right of the subplots.} 
\label{fig:resolution filfinder}
\end{figure*}

\begin{figure*}
\includegraphics[width=\textwidth]{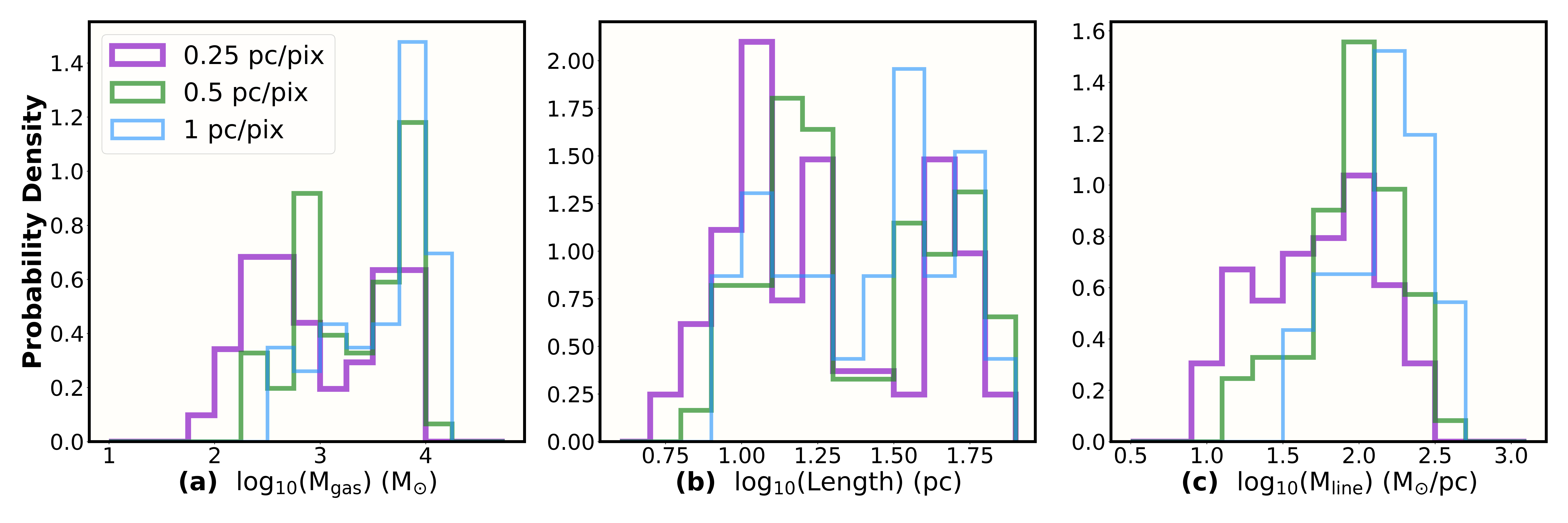}

\caption{The Probability Density Function (PDF) of filament properties. The three panels show PDFs of filament properties for three different spatial resolutions (Purple: 0.25pc/pix, Green: 0.5pc/pix, Blue: 1pc/pix). Left panel: PDF of filament mass derived from the column density maps. Middle panel: PDF of filament length. Right panel: PDF of line mass. The PDF shows that the distribution of filament properties varies with spatial resolution. The higher resolution shows more filaments with smaller masses, shorter lengths, and lower line masses. The distribution in mass, length, and line mass suggests they are different populations of filaments.}
\label{fig:pdf}
\end{figure*}

In this work, we identify filaments across the three regions at three distinct resolutions. However, the number of filaments identified varied depending on the resolution and the specific snapshot in time. Initially, we find a total of 189 filaments across all snapshots and resolutions combined. The filament mass $M\rm_{fil}$, length $L\rm_{fil}$, and the stellar mass $M_\mathrm{{stellar}}$ of these filaments are derived using the method mentioned in Section \ref{sec:calculation}. The line mass (mass per unit length) of the filaments is defined as $m = M_\mathrm{{fil}}/L_\mathrm{{fil}}$.

One of the chief findings of \cite{hacar2022initial} is that filament structures are hierarchical, with nested smaller sub-structures contained within larger filaments. Thus, a filament that appears to be a single object may reveal smaller substructures when observed at higher resolution. Figure \ref{fig:resolution filfinder} shows the identified filament skeletons in Region A at three different resolutions for snapshot 3. As expected, for the higher resolution, more filaments are identified, while for the lowest resolution, only two filaments are identified. The filaments seen at higher resolution are not nested within the structures identified at lower resolution but rather are new objects that have not been previously identified. At lower resolutions, these new objects cannot be identified because their morphology remains unresolved, causing them to appear as 'blobs' with a skeleton length shorter than the threshold. Specifically, we find 82 filaments at a resolution of 0.25 pc/pix in all snapshots across the three regions, 61 filaments at 0.5 pc/pix, and 46 filaments at 1 pc/pix.

The mass Probability Distribution Functions (PDF) of the filaments identified at different resolutions are shown in Figure \ref{fig:pdf}. Panel (a) indicates that there is a different distribution of mass among the filaments at different resolutions. For example, the 0.25 pc/pix resolution filaments exhibit numerous low-mass filaments, with two peaks at $\sim 200 \ \rm M_{\sun}$ and $\sim \  5000 \ \rm  M_{\sun}$, whereas the 1 pc/pix resolution filaments have a mass distribution with only one peak at $\sim 7000 \ \rm M_{\sun}$. Meanwhile, the length PDF shown in Figure \ref{fig:pdf} (b) reveals that the \textsc{FilFinder} filaments distribute from $\sim$ 8 pc up to $\sim$ 90 pc, with 0.25 pc/pix resolution filaments exhibiting two peaks at $10^{1.7}$ pc ($\sim$ 60 pc) and at $10^{1.1}$ pc ($\sim$ 12 pc). 0.5 pc/pix and 1 pc/pix resolution filaments also exhibit a two-peak distribution with similar peak values. However, as the resolution becomes lower, the proportion of longer filaments increases while the proportion of shorter filaments decreases. The mass and length of filaments display a distinct bimodal distribution in the PDF, without any indications of hierarchy. Hierarchical structures should manifest as self-similar patterns characterised by a power-law PDF. This might be due to the resolution variation being insufficient to show the nested structures and the sample numbers being inadequate. The bimodal distribution is due to the limitations of our sample, which only showcases typical samples that the algorithm can identify within these three regions at specific resolutions.

The line mass of \textsc{FilFinder} filaments varies over a large range, from $\sim 5$ to $\sim 500 \ \rm M_{\sun}\ pc^{-1}$. Filaments with 0.25 pc/pix and 0.5 pc/pix resolutions have a peak at $\sim 100 \ \rm M_{\sun}\ pc^{-1}$, while those with 1 pc/pix resolution have a higher peak at $\sim 200 \ \rm M_{\sun}\ pc^{-1}$. Unlike the other quantities, the line mass distribution does not show a clear bimodal profile, indicating a narrower range of line mass. This may suggest that underlying mechanisms are constraining the distribution.

\subsection{Mass-Length relation}
\label{sec:M-L relation}

\begin{table}{
\caption{Tracked-filaments in the three regions and their properties for a resolution of 0.25 pc/pix.}
\begin{threeparttable}

\begin{tabular}{cccccc}
\hline
id   & Snapshot & Mass        & Length & $m $             & M$\rm _{Stellar}$ \\
     &          & M$_{\odot}$ & pc     & M$_{\odot}/ \rm pc$ & M$_{\odot}$ \\ 
(1) & (2) & (3)     & (4)           & (5)            & (6)             \\ \hline
RAF1 & t3       & 544         & 14     & 38                  & 0           \\
     & t8       & 721         & 13     & 55                  & 0           \\
     & t18      & 1024        & 18     & 56                  & 135         \\
     & t23      & 1465        & 22     & 66                  & 289         \\
RAF2 & t3       & 4287        & 38     & 114                 & 64          \\
     & t8       & 5860        & 50     & 118                 & 542         \\
     & t13      & 4790        & 43     & 112                 & 596         \\
     & t18      & 5072        & 45     & 112                 & 643         \\
     & t23      & 9899        & 78     & 127                 & 692         \\
RAF3 & t3       & 2771        & 43     & 65                  & 0           \\
     & t8       & 4459        & 50     & 89                  & 58          \\
     & t13      & 4450        & 40     & 111                 & 193         \\
     & t18      & 2296        & 40     & 57                  & 129         \\
RBF1 & t3       & 3760        & 59     & 64                  & 643         \\
     & t5       & 4426        & 59     & 76                  & 1073        \\
     & t7       & 4714        & 58     & 81                  & 1609        \\
     & t9       & 5739        & 65     & 88                  & 2032        \\
     & t11      & 6089        & 57     & 107                 & 2333        \\
     & t13      & 5723        & 49     & 116                 & 2608        \\
     & t15      & 5687        & 32     & 177                 & 2801        \\
RCF1 & t3       & 8556        & 43     & 198                 & 2895        \\
     & t4       & 8988        & 48     & 188                 & 3309        \\
     & t5       & 8597        & 41     & 208                 & 3179        \\
     & t6       & 8820        & 42     & 210                 & 3188        \\
RCF2 & t3       & 4653        & 58     & 80                  & 59          \\
     & t4       & 5986        & 62     & 96                  & 873         \\
     & t5       & 7177        & 63     & 114                 & 1687        \\
     & t6       & 9128        & 59     & 154                 & 3034        \\
RCF3 & t3       & 2309        & 16     & 144                 & 930         \\
     & t4       & 2465        & 16     & 153                 & 1072        \\
     & t5       & 2623        & 17     & 154                 & 1188        \\
     & t6       & 2796        & 20     & 143                 & 1414        \\
RCF4 & t3       & 919         & 9      & 102                 & 359         \\
     & t4       & 997         & 9      & 111                 & 379        \\
     & t5       & 819         & 7      & 117                 & 300        \\

RCF5 & t3       & 5362        & 31     & 173                 & 2128        \\
     & t4       & 4612        & 19     & 245                 & 2395        \\
     & t5       & 4852        & 18     & 264                 & 2472        \\
     & t6       & 5328        & 26     & 201                 & 2770        \\ \hline
\multicolumn{6}{l}{}\\

\end{tabular}
\begin{tablenotes}
\item[] \textbf{Note. } (1) The ID number of the filament (2) Time snapshot (3) The total gas mass, including the gas portion within the sink mass (4) The length of the skeleton identified by \textsc{FilFinder} (5) The line mass (6) The mass of the stellar portion of the sinks, which is 33\% of the sink mass.
\end{tablenotes}
\label{tab:filaments properties}
\end{threeparttable}}
\end{table}

\begin{figure*}
\includegraphics[width=\textwidth]{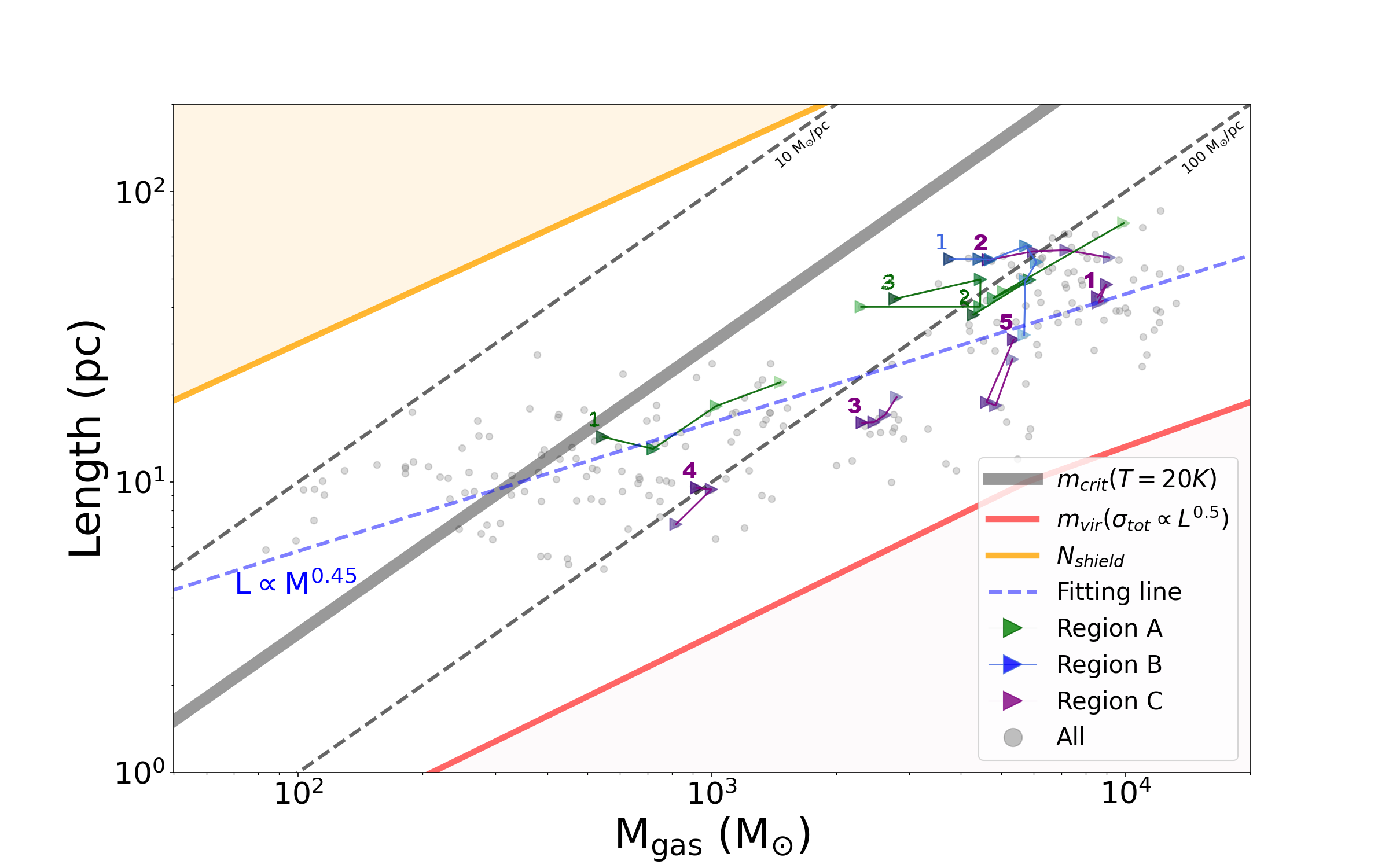}

\caption{The mass-length correlations of our simulated filaments are observed in all regions across all snapshots. The colours of the evolutionary paths connecting different snapshots of a filament represent different regions: green for Region A, blue for Region B, and red for Region C. The colour of the markers indicates the order of time evolution, starting with dark and ending with light. The grey circles in the background represent the entire filament population. The number refers to the ID of the filament and marks the position of the first snapshot in each time series, which is connected with a line. The grey solid line corresponds to the critical line mass of a filament derived from a hydrostatic isothermal cylinder model (see Equation \ref{equ:mcrit}). The red solid line shows $m_{\mathrm{vir}}$ including turbulent motions (see Equation \ref{equ:mvir}). The yellow solid line corresponds to the column density $N_{\rm shield}$ required for self-shielding (see Equation \ref{equ:Nshield}). Almost all samples lie between the yellow and red lines, corresponding to the upper and lower limits. The blue dashed line represents the scaling relation $L \propto M^{0.45}$ fitted to our samples. The evolved filament paths show a tendency to shift rightward, indicating overall accretion.}

\label{fig:M-L all}
\end{figure*}

Figure \ref{fig:M-L all} presents the mass and length properties of all our samples (including different resolutions) in log-space.

We follow the approach of \cite{hacar2022initial} to compare the distribution of our filament sample to equilibrium filaments. We first compare our results with the critical line mass, which is analogous to the isothermal Jeans mass of molecular clouds \citep[e.g.][]{klessen2000formation,klessen2001formation,larson2005thermal}.
The gray solid line in Figure \ref{fig:M-L all} represents the critical line mass of a hydrostatic, isothermal cylinder \citep{stodolkiewicz1963gravitational,ostriker1964equilibrium}:

\begin{equation}
\label{equ:mcrit}
m_{\mathrm{crit}}(T)= \frac{2c_{s}^2}{G} \sim 16.6 (\frac{T}{\mathrm{10K}}) \  \rm M_{\sun}\ pc^{-1} 
\end{equation}

where $c_s$ is the sound speed, which is given by

\begin{equation}
c_s=\sqrt{\frac{k_B T}{\mu m_p}}
\end{equation}

Here $k_B$ is the Boltzmann constant, $m_p$ is the proton mass, the molecular weight $\mu = 2.8$, and $T$ is the gas kinetic temperature.
Filaments with a line mass $m > m_{\mathrm{crit}}$ (supercritical) are unstable and undergo collapse due to their own gravity, while those with $m < m_{\mathrm{crit}}$ (subcritical) can remain in hydrostatic equilibrium.

The red solid line represents the virial line mass $m_{\mathrm{vir}}$, which is given by:

\begin{equation}
\label{equ:mvir}
m_{\mathrm{vir}}= \frac{M}{L} \simeq \frac{2c_{s}^2}{G} (1 + \frac{L}{0.5 \,\rm pc})
\end{equation}

It is important to note that this virial mass includes an extra non-thermal pressure contribution based on the emission line-width scaling seen in Milky Way filaments (see Figure 2 in \cite{hacar2022initial}). The thermal-only case is shown by the grey line, and thus, between the grey and red lines, the filament is only stable if the observed supersonic velocity dispersion is interpreted as a supportive pressure force. Below the red line, gravity dominates, resulting in collapse and fragmentation. Above this line, a combination of thermal and non-thermal pressure dominates, potentially attributable to turbulence.

In our case, as shown in Figure \ref{fig:M-L all}, most of the filaments are distributed below the thermally critical line mass.This might be because the model for thermally critical line mass is based on several simple assumptions, including that the filament is in hydrostatic equilibrium, isolated, and isothermally. However, filaments are generally more complex than the ideal model, which can result in differences from the theoretical predictions.

The yellow solid line indicates an upper limit for the filament distribution to remain molecular, driven by self-shielding against UV radiation. Its relation is derived from \citep{hacar2022initial}:

\begin{equation}
\label{equ:Nshield}
L \simeq 1.5(\frac{M}{\mathrm{M_{\sun}}})^{0.65} (\frac{N_{\mathrm{shield}}}{10^{21}\ \rm cm^{-2}})^{-0.65}\ \rm pc
\end{equation}

where $N_{\rm shield}=10^{21}\rm cm^{-2}$ \citep{van1988photodissociation}. It is based on the assumption that the central density is 10 times the average density and the filament is in hydrostatic equilibrium. However, the upper limit relation is only a rough estimate because materials in the surrounding environment also contribute to the shielding effect.

Most of the filaments in our sample lie in the regime between the yellow and red lines (with only one exception, which is slightly below the virial mass line). This is consistent with the observed distribution of filaments in the Milky Way \citep{hacar2022initial}, and suggests that the distribution of our simulated filaments aligns well with the expectations from these simple models. However, this does not imply that the filaments are static objects because they can still grow (by moving rightward), shrink (by moving leftward), and fragment to form stars as the evolutionary paths shown in Figure \ref{fig:M-L all}. We will discuss this evolutionary path in more detail in Section \ref{sec:M-L evolution}.

\subsection{Evolution of Filament Properties}
\label{sec:M-L evolution}

\cite{hacar2022initial} show that molecular filaments follow a continuous distribution across a large range of mass and lengths, corresponding to an approximate scaling relation $L \propto M^{0.5}$. Over this range, filaments possess different stability and dynamical properties. For example, filaments that exceed the expected $m_{\mathrm{vir}}$ for a hydrostatic filament would be more susceptible to collapse and fragmentation. On the other hand, many of the shorter filaments in \citet{hacar2022initial} display a sub-critical line mass, suggesting these filaments might be transient structures or evolving \citep{hacar2022initial}. In this case, the distribution of filaments in the M-L diagram might be able to trace the evolving status of filaments.

Here, we use the highest resolution (0.25 pc/pix) when tracking the evolutionary path, as this identifies the greatest number of filaments.
We examine all of these filaments and select only those that were clearly identifiable in at least three snapshots for our subsequent analysis of their evolution, which we refer to as tracked-filaments.
This is because the small structure undergoes significant morphology changes during evolution, particularly in complex surrounding gas environments, which make it difficult to track with an automatic algorithm.
The properties of these tracked-filaments are listed in Table \ref{tab:filaments properties}. The ID number of the tracked-filaments corresponds to the region they belong to. For example, 'RBF1' stands for Region B Filament 1. We finally track three filaments in Region A, one filament in Region B, and five filaments in Region C. Not all tracked-filaments can be identified in every snapshot because of the changing morphology. RAF2 and RCF4 are lost from our tracking before the final snapshot. RAF1 is the only filament that is lost and re-found due to the algorithm bias.
After we trace the time evolution of filaments, we can now test how filaments evolve in the M-L diagram.

\subsubsection{Accretion}
\label{sec:Accretion}

Figure \ref{fig:M-L all} shows that 7 of 9 filaments undergo a pronounced accretion process as they move from left to right on the M-L plot. This is the dominant evolutionary path for all the filaments in our samples.
This process can result in either an increase in both mass and length or a significant increase in mass with little change in length. In both scenarios, the filamentary structures accrete material from their surroundings. Filaments RAF1, RBF1, RCF2, and RCF3 all exhibit a significant accretion path with continuous mass increases.

We consider the filament growth rate by comparing the line mass of two adjacent snapshots to assess the rate of line mass increase. The equation employed for calculating the growth rate is presented below:

\begin{equation}
\Dot{m} = \frac{\Delta (m) }{\Delta t}
\label{equ:growth rate}
\end{equation}
where $m$ is the line mass of the filament, $m=M/L$, while $M$ refers to the mass of the filament and $L$ represents its length. The time between the two snapshots is denoted by $\Delta$ t with a unit of Myr.  A positive filament growth rate might be due to accretion or length shortening. We remove the negative values caused by segmentation. This is because the line mass of filaments is not uniform, so it might lead to a decrease in line mass after segmentation. Since we are only interested in studying the relationship between accretion and line mass, such drastic changes are not considered.

\begin{figure}
\includegraphics[width=0.5\textwidth]{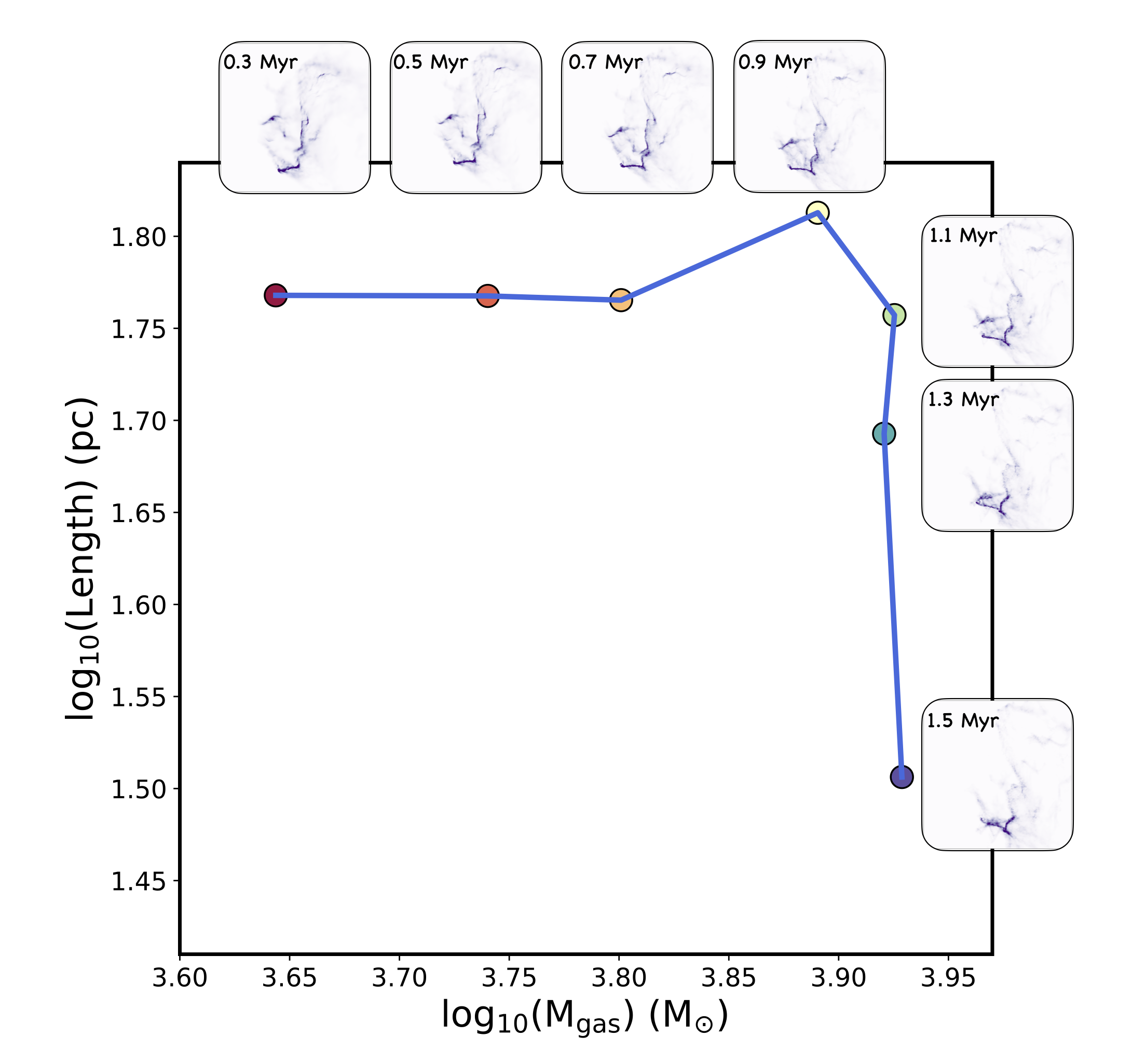}

\caption{The evolution of the mass-length correlation of filament RBF1. The surrounding subplots display the $\rm H_2$ column density maps of each snapshot, along with their corresponding evolution times indicated in the top-left of the subplots. Filament RBF1 experiences an accretion at the beginning and then undergoes a dispersal process.}
\label{fig:M-L RE}
\end{figure}

RBF1 is the longest-tracked filament in our samples, and it is relatively independent, simple, and at an early stage of evolution with little star formation. Therefore, we chose RBF1 for further study to analyse the long-term evolution process of the filament (10$^6$ years). Figure \ref{fig:M-L RE} shows a simplified version of the M-L plot containing only RBF1. As shown in Table \ref{tab:filaments properties}, RBF1 increases in mass and length in the earlier part of the evolution and then drops in the later part. Meanwhile, the line mass increases the whole time.

\begin{figure*}
\includegraphics[width=\textwidth]{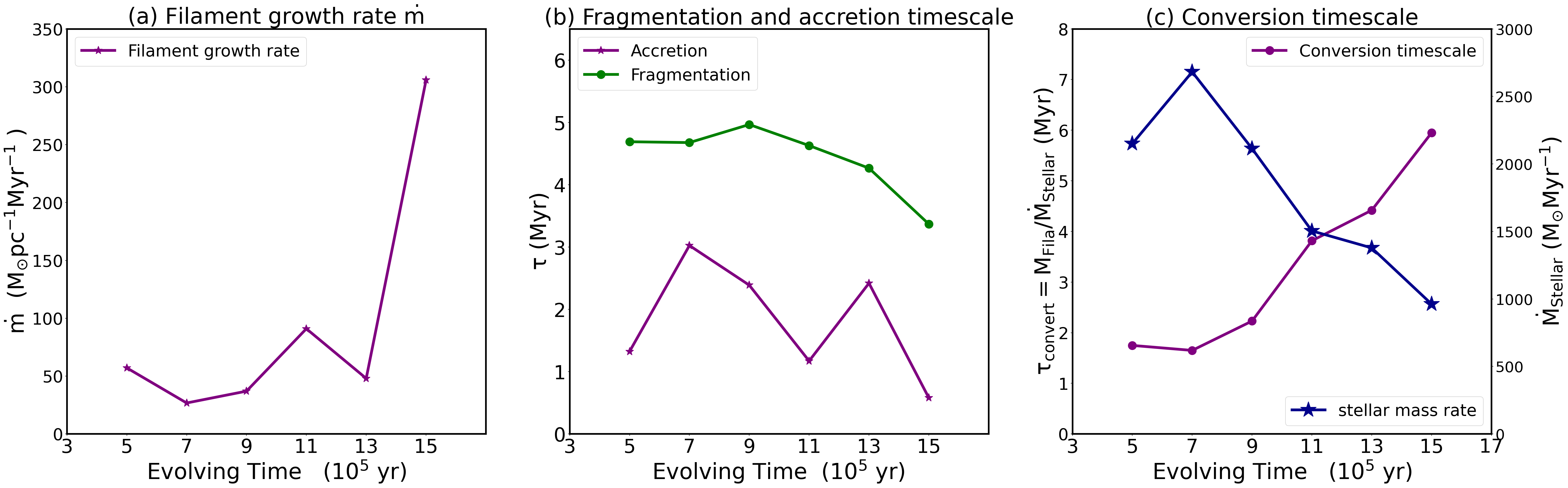}

\caption{Evolutionary trends of the filament RBF1 over time.
\textbf{(a)} The filament growth rate, \( \dot{m} \), illustrates the rate at which the line mass of the filament grows.
\textbf{(b)} Timescales for accretion and fragmentation.
\textbf{(c)} Conversion timescale, \( \tau_{\text{convert}} \), and stellar mass rate, \( \dot{M}_{\text{Stellar}} \). The conversion timescale denotes the timescale at which gas in the filament is converted into stars. The stellar mass rate indicates the rate at which stars form in the filament.
The calculations for these plots are described in Section \ref{sec:M-L evolution}. The growth rate exhibits an increasing trend, and $\tau_{\mathrm{frag}}$ is larger than $\tau_{\mathrm{acc}}$ during the evolution period. The conversion timescale is increasing while the stellar mass rate drops.
}
\label{fig:timescale}
\end{figure*}

In Figure \ref{fig:timescale} (a), we present the evolution of the filament growth rate across the snapshots. For RBF1, the mean filament growth rate in dust is 94 M$_{\odot}\  \rm pc^{-1}\  Myr^{-1}$. The filament growth rate hovers around a few dozen before the last snapshot and then increases to 300 $\rm M_{\odot}\  \rm pc^{-1}\  Myr^{-1}$ in the final snapshot. However, this final increase is driven more by a decrease in the length of the filament, as we will discuss in the next section.

\subsubsection{Segmentation}
\label{sec:Segmentation}
\begin{figure*}
\includegraphics[width=\textwidth]{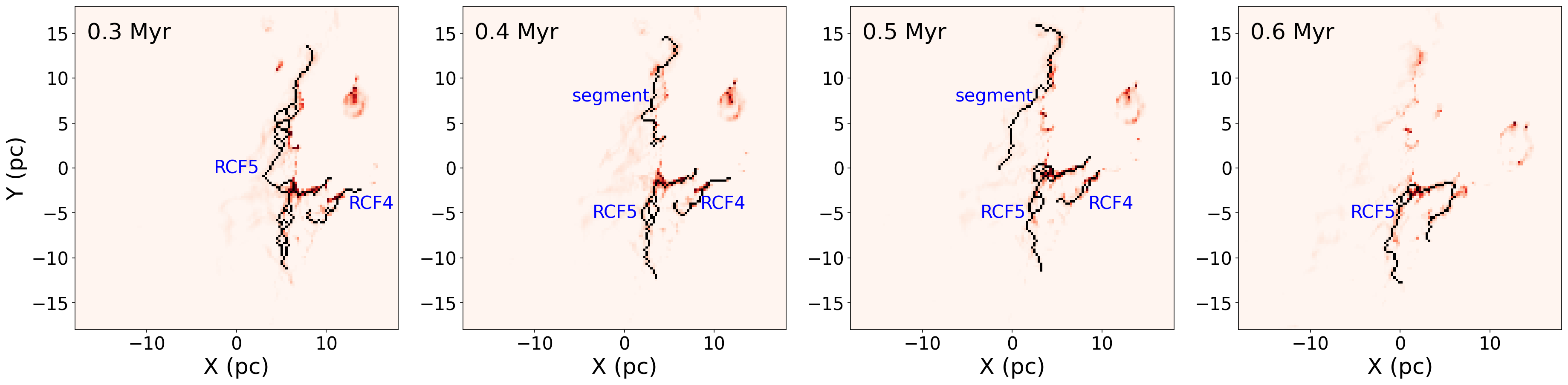}

\caption{The evolution of filament RCF5. The black dashed line is the filament structure, including the skeleton and branches identified by \textsc{FilFinder}. The background red contour is the H$_2$ column density map. The label in the top-left corner of each panel shows the corresponding snapshot. Filament RCF5 segments into two parts in snapshot 4, and merges with filament RCF4 by the algorithm in the last snapshot 6. The segment from RCF5 becomes more split and cannot be identified in snapshot 6. } 
\label{fig:RCF5}
\end{figure*}

The mass and length of the filaments do not uniformly increase at all times. Fragmentation often occurs when star formation begins during the latter stages of filament development, causing the filament to segment into multiple smaller pieces. RCF5 is one of a filament sample that undergoes such a segmentation process and then resumes growth, as is shown in Figure \ref{fig:RCF5}.

At snapshot 4, the filament RCF5 breaks into two parts and combines with RCF4 at snapshot 6. These processes lead to a 'U'-turn trend in the M-L diagram (see the purple path 5 in Figure \ref{fig:M-L all}). The segmentation of RCF5 splits the filament into an upper and bottom part, with the upper part having a mass of approximately $900\ \rm M_{\odot}$ and a length of 16 pc, while the bottom part is the main part with a mass of approximately $4600\ \rm M_{\odot}$ and a length of 18 pc.
When gas is consumed by star formation in the filament, gaps appear along its length, and it is broken into smaller pieces by the filament-finding algorithm. In the last snapshot, the upper segment splits from the bottom and is no longer identified as a single structure by the algorithm. During this process, the gas gathers into dense cores, eventually transforming into stellar mass.

The skeleton determined by the algorithm is sensitive to the column density threshold in these regions. This sensitivity can significantly alter the skeleton path since the algorithm draws the path based on the surrounding pixels. For instance, the choice of threshold is likely responsible for the segmentation and merging of RCF5.

In the M-L diagram, filaments tend to move to the right due to accretion until they are disrupted by segmentation (shift towards the bottom-left), and then they resume moving rightward. 
For example, although RAF3 and RCF5 show a significant decrease in mass in the last snapshot, it is probably due to the segmentation mechanism changing the structure of the filament identified rather than a rapid change in the gas density.

As suggested by \citet{hacar2022initial}, the balance between fragmentation and accretion also determines whether a filament can survive. If fragmentation is too rapid, all the gas in the filament will be in the fragments rather than a continuous density structure, and so a filament will not be identified. Figure \ref{fig:timescale} panels (b) show the accretion and fragmentation time of filament RBF1 calculated using the below relations \citep{hacar2022initial}:

\begin{equation}
\tau_{\mathrm{acc}} = \frac{m}{\Dot{m}}
\end{equation}

\begin{equation}
\tau_{\mathrm{frag}} = 0.5 \cdot (\frac{L}{\mathrm{(pc)}})^{0.55} \rm  
\end{equation}

The mean accretion timescale is 1.8 Myrs, similar to that found in the Galactic Cold Core Herschel sample (1--2 Myrs, \citealt{rivera2017galactic}), while the mean fragmentation timescale is 2.5 Myrs. The fragmentation timescale $\tau_{\mathrm{frag}}$ is consistently larger than $\tau_{\mathrm{acc}}$ during the evolution period for Filament RBF1. The longer timescale of fragmentation shows that filament RBF1 more rapidly accretes gas from its surroundings than it can fragment into smaller structures. This is also shown by the increasing line mass. This accounts for why filament RBF1 can be followed for over 1 Myr in our simulations, as the gas is replenished as the filament evolves despite the ongoing star formation.

Figure \ref{fig:timescale} panel (c) presents the conversion timescale for all the gas in the filament gas to be transformed into stars, derived by:

\begin{equation}
t_{\mathrm{convert}}= \frac{M_{\mathrm{Fil}}}{\dot{M}_{\mathrm{Stellar}} }
\end{equation}

where $\dot{M}_{\mathrm{Stellar}} = \Delta {M}_{\mathrm{Stellar}} / \Delta t $ is the stellar mass conversion rate. The conversion timescale increases from $\thicksim $ 1.7 Myrs to $\thicksim $ 6 Myrs, with an average value of 3.3 Myrs. Compared to the mean accretion timescale (1.8 Myrs), the conversion timescale is similar at the early stage of the evolution and much larger at the end. The mass conversion rate $\dot{M}_{\mathrm{Stellar}}$ also supports this conclusion, as it decreases from 2800 $\rm M{\odot}\ Myr^{-1}$ to 1000 $\rm M_{\odot}\ Myr^{-1}$ over the evolution, representing a decline of nearly 2/3.
In this specific case, the reduction in star formation is due to the densest part of the filament being converted into stars and the remainder of the filament being less dense and, consequently, less actively star forming.

\subsubsection{Dispersal}
\label{sec:Dispersal}

The final mechanism we see is 'dispersal', where rather than the filament being consumed by internal star formation, it is disrupted due to external forces.
Sources of external pressure may come from mechanical and radiation feedback or compression by the galactic spiral potential, acting at different scales. When an external force interacts with a filament, it may undergo a morphological change resulting in a shorter length due to a combination of compression of the dense gas and removal of the more diffuse gas.
This is exemplified well by Filament RBF1, which, as seen in Figure \ref{fig:M-L RE}, undergoes a sharp decrease in length at the end of its evolution while retaining most of its mass.
Alternatively, filaments may also be stretched by shear \citet{smith2016nature} or by differential rotation \citep{smith2014nature,duarte2017evolution}.
The surrounding subplots in Figure \ref{fig:M-L RE} show the morphology change of REF1, with the contour of $\rm H_2$ column density. It can be clearly seen that the upper gas structure is moving downward, causing a dispersal with length shortening.

\begin{figure}
\includegraphics[width=0.5\textwidth]{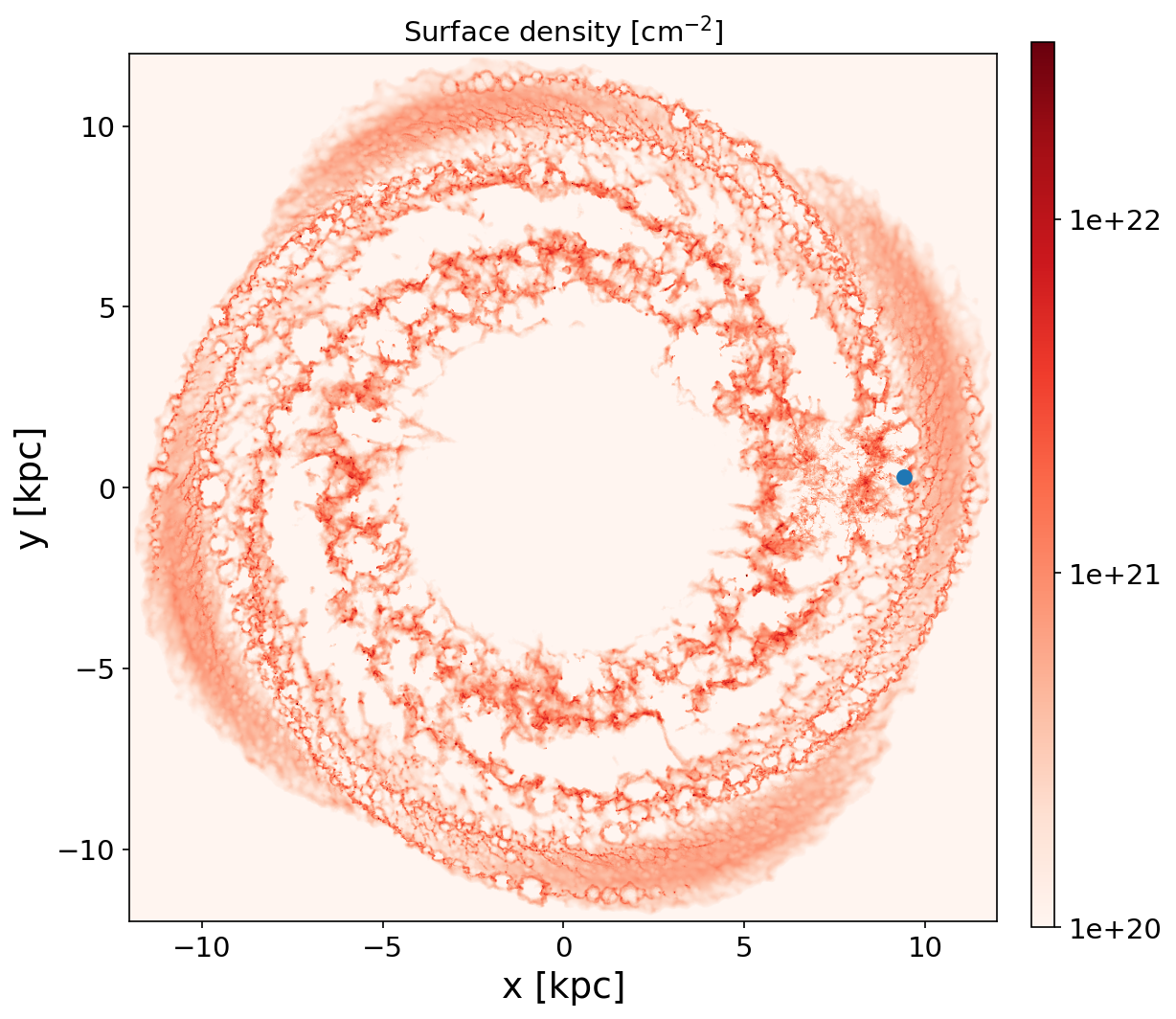}

\caption{Illustration of the location of RBF1. The blue dot is the position of RBF1, and the background red contour is the surface density face-on projection map of the feedback-dominated galaxy from the Cloud Factory. RBF1 is located in the inter-arm region.} 
\label{fig:RB illustrate}
\end{figure}

To determine the reason for this, we first inspect the location of RBF1 in the galaxy. RBF1 is located in between two spiral arms, as shown in Figure \ref{fig:RB illustrate}. We then examine the supernovae locations from the Cloud Factory model. However, we find no supernova explosion within a distance of 50 pc from RBF1 during the time frame under consideration. Considering RBF1 is not in the spiral arm, we suggest the dispersal of RBF1 is probably due to the differential rotation of gas moving between the arms. Such dispersal can be visually observed in Figure \ref{fig:cloud nh2 maps}. In the last snapshot 15, the bottom part of RBF1, which shows active star formation, appears denser than in the first snapshot 3. This indicates that diffuse gas from the upper part has moved to the bottom part during this time period.

However, we only find one filament associated with this mechanism. This is likely due to the fact that only Region B is located in the interarm region, and we stop tracing the evolution before many possible supernova explosions could occur after active star formation.

\subsubsection{Summary}
\label{sec:evo mechanism}

\begin{figure}
\includegraphics[width=0.5\textwidth]{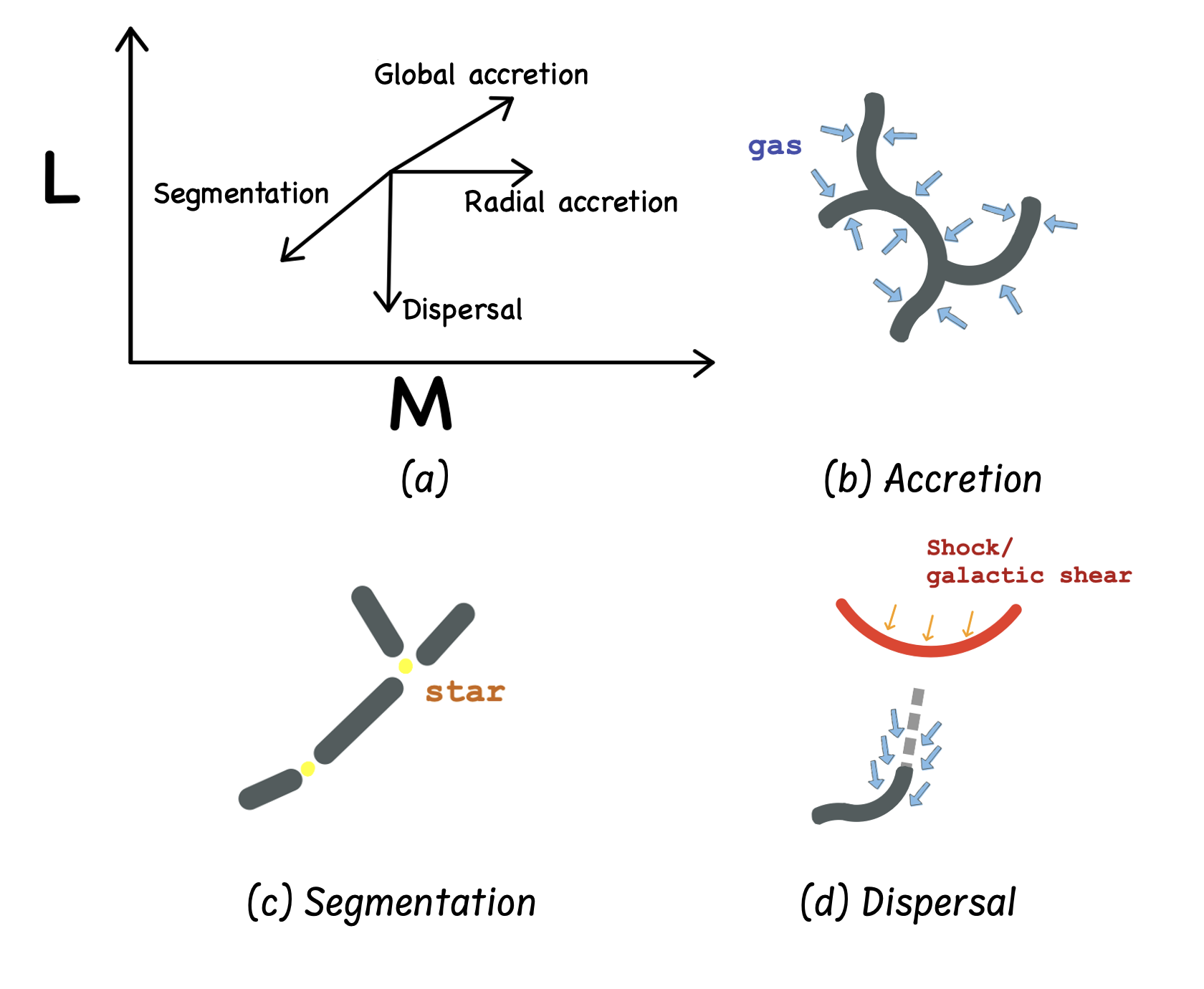}

\caption{Illustration of the mechanisms of filament evolution.
(a) The arrows show how the position in the M-L diagram changes under different mechanisms.
(b) Accretion: Gas is accreted onto the spine.
(c) Segmentation: Filaments undergo segmentation due to star formation within them.
(d) Dispersal: External forces impact the filament, causing a change in morphology.} 
\label{fig:illustrate}
\end{figure}

We therefore propose that the evolutionary progression of filaments in our models in the M-L phase diagram can be reduced to three physical processes, which are illustrated in Figure \ref{fig:illustrate}. Figure \ref{fig:illustrate} (a) shows the directionality of the evolutionary path, while the corresponding four physical mechanisms are displayed in the sub-figures below in (a). Figure \ref{fig:illustrate} (b) displays accretion, which is characterised by an increase in filament mass over time, while the length may either increase or remain steady. The entire gas filamentary structure is undergoing accretion from its surrounding material, leading to significant increases in mass.

The remaining two mechanisms result in a decrease in mass and length.
Segmentation of filaments due to the conversion of gas into stars, as shown in Figure \ref{fig:illustrate} (c), causes both a decrease in filament mass and length. Meanwhile, Figure \ref{fig:illustrate} (d) illustrates dispersal, characterised by a significant decrease in length with little change in mass. This mechanism is attributable to external forces acting on the filament, such as shocks generated by supernovae or shear resulting from larger-scale galactic motions.

In our samples, 7 out of 9 tracked-filaments exhibit significant accretion paths, except RCF1 and RCF4. Both RCF1 and RCF4 are active star forming regions and have only evolved for a short period of time. Three filaments (RAF3, RCF4, and RCF5) experience a segmentation process, while only one filament, REF1, undergoes a dispersal process. These results indicate that accretion is the dominant mechanism in filament evolution. However, it is important to note that these mechanisms are merely observed in our samples and do not represent all possible scenarios. All these mechanisms control the distribution of filaments in the M-L diagram, maintaining the hierarchical nature of filaments in a large spread range of scale.

We want to emphasise that we do not incorporate internal dispersal mechanisms in our study because our simulation lacks early stellar feedback components such as jets, outflows, and photoionization. The absence of these factors in our paper should not be interpreted as an indication that filament dispersal through these mechanisms is unimportant; it simply means that we have not included them in our current models. 
We plan to investigate these internal processes in future research.

\section{Discussion}

\subsection{The prominent mechanism in filament evolution: Accretion}
\label{sec:accretion disscusion}

In section \ref{sec:M-L evolution}, a rightward shift in the Mass-Length diagram for most filaments pointed to accretion as a probable cause. To validate this, we first focus on the line mass $m$, a parameter sensitive to accretion activity.

\begin{figure}
\includegraphics[width=0.5\textwidth]{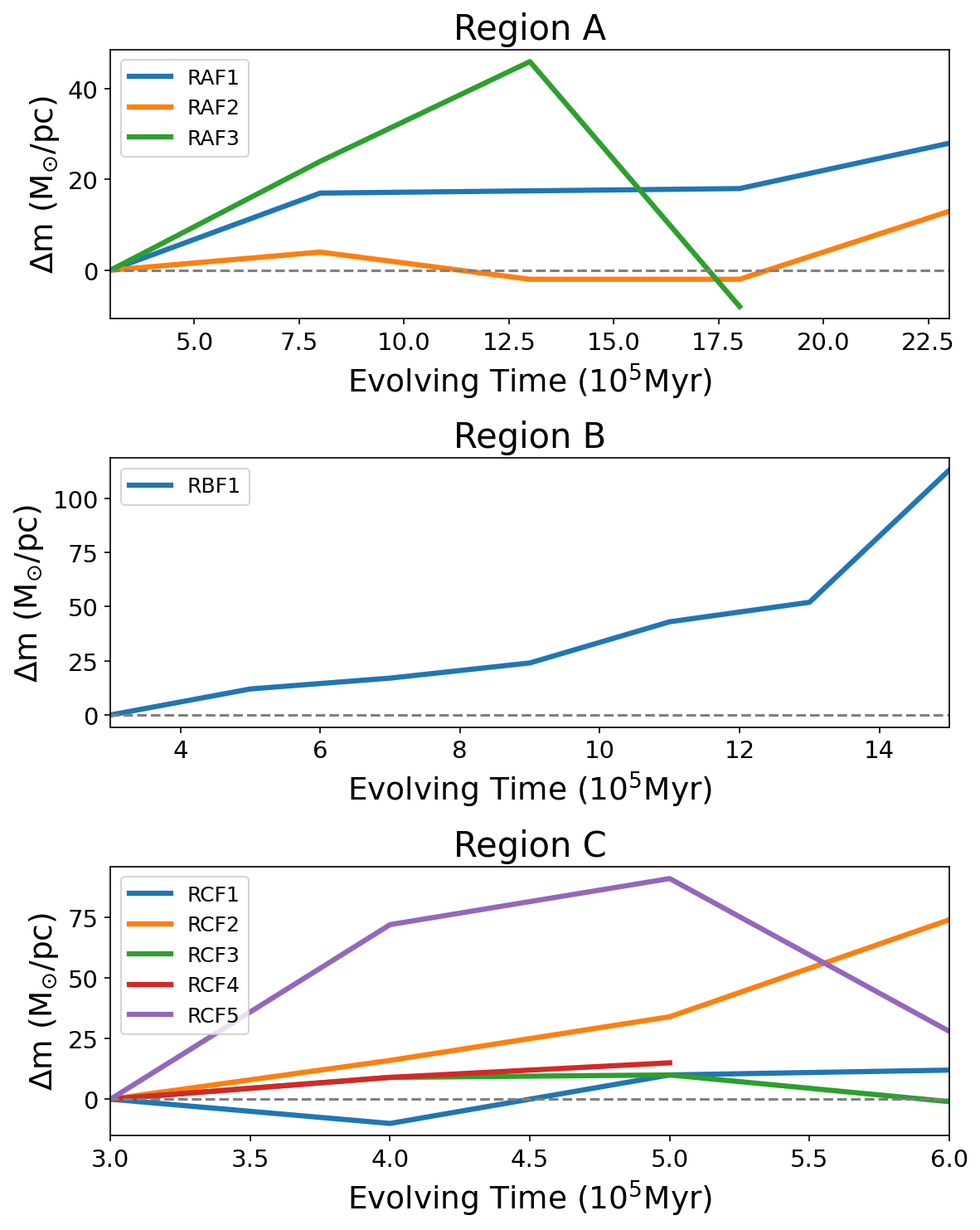}

\caption{The change in line mass of all tracked filaments is measured by comparing it with their first snapshot: $\Delta m = m_i - m_0$, where $m_0$ represents the line mass of the initial snapshot and $m_i$ represents the line mass of subsequent snapshots. It is observed that the majority of filaments experience an increase in line mass.} 
\label{fig:line mass}
\end{figure}

\cite{chira2018fragmentation} find that the average line masses of filaments in their simulation always increase with time. We also notice the same trend as shown in Figure \ref{fig:line mass}. Most of our tracked-filaments samples (7 of 9) end up with a higher line mass compared to their first snapshot. The increasing line mass is reflected in the M-L diagram as filaments tend to shift towards the right, as shown in Figure \ref{fig:M-L all}. In the maximum case, the line mass can increase by over 100 $\rm M_{\odot} \rm \ pc^{-1} \ Myr^{-1}$ compared to the initial line mass (RBF1).

However, the observed increase in line mass may also be influenced by algorithmic identification, as structural variations in the same filament across different snapshots could affect the line mass calculations. To eliminate this potential confounding factor, an analysis of the raw simulation data was conducted for the filaments.

In real observation, the kinematics of molecular gas can be inferred through the Local Standard of Rest (LSR) velocity $V_{\text{LSR}}$. For example, a velocity gradient perpendicular to the filament may suggest rotational motion or gravitational infall within the filament. Alternatively, it might indicate the projection effects of an inclined planar structure or the parallel overlapping of multiple narrow filaments \citep{storm2014carma,fernandez2014carma}.

\begin{figure*}
\centering
\includegraphics[width=1\linewidth] {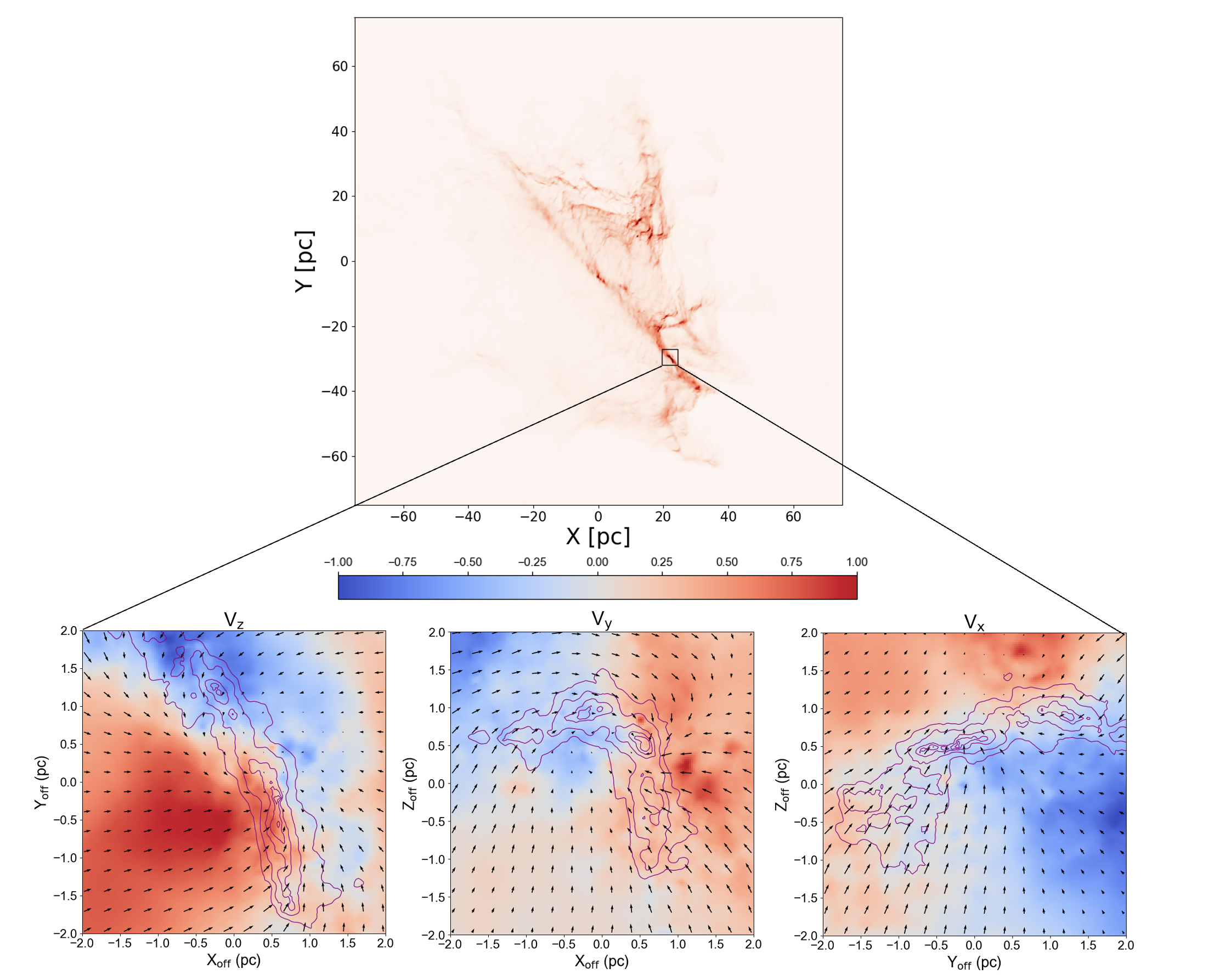}
\caption{The figure provides a detailed visualisation of a filament spine and its associated velocity components. In the top panel, the column density map of Region A at snapshot 3 is displayed, with a designated rectangle highlighting the area expanded upon in the bottom panels. The bottom panels present the velocity structures from the Cloud Factory simulation raw data within the selected region. The left, centre, and right panels represent the XY, XZ, and YZ plane projections, respectively. The background colormap shows the standardised velocity components of \( V_z \), \( V_y \), and \( V_x \). The colorbar representing the standardisation velocity component is displayed in the centre of the picture. The purple contours indicate the dense gas structures. The black arrows represent the velocity vector in its respective projection. All three bottom panels show an ongoing accretion process.}
\label{fig:accretion velocity}
\end{figure*}

In Figure \ref{fig:accretion velocity}, we present the velocity structure of a segment of the spine of a filament in Region A snapshot 3 to further inspect the ongoing accretion process within the filament. Across the three different projected view angles, the velocity components exhibit symmetrical distribution, indicating gas convergence towards the center of the filament from all directions. It should be noted that the velocities used here are derived directly from simulation raw data, and thus cannot be simply compared with $V_{\text{LSR}}$ from real observations. 

In addition, although there is a broadly symmetrical distribution of velocities, there are also complex patterns present. The XZ plane projection reveals that the filament is not an ideal cylinder object, but instead exhibits a curvature. The projected velocity components (shown by black vectors) do not always follow a perpendicular trend to the filament spine, but might also be parallel to it. In fact, the velocity vectors appear to be oriented towards the densest region. This indicates that the gas accretion is shifting towards the dense core, either from the surroundings or along the filament itself.

\begin{figure}
\centering
\includegraphics[width=1\linewidth]{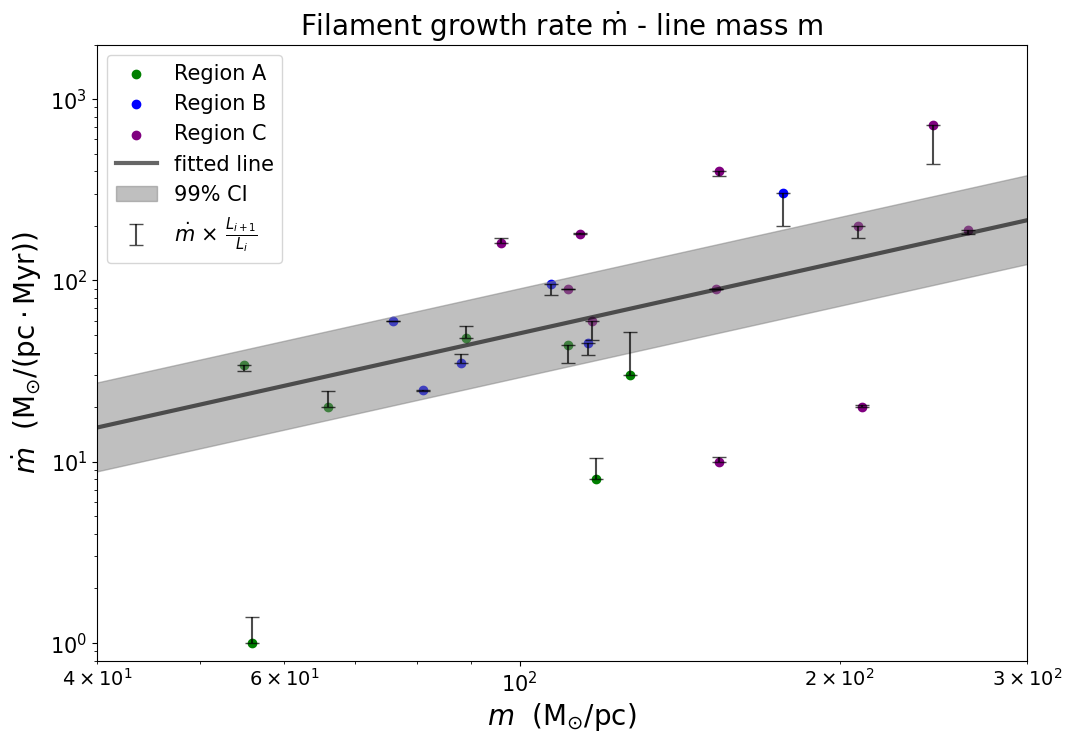}
\caption{The scatter plot illustrates the relationship between the filament growth rate, $\dot{m}=m_{i+1}-m_i$, and the line mass, \(m\), for three regions: Region A (green points), Region B (blue points), and Region C (purple points). The bar corresponding to $\dot{m}$ multiplied by a length correction factor $\frac{L_{m+1}}{L}$. The fitted line shows a potential relationship between the growth rates and the line mass, indicating these filaments might be gravity-dominated. }
\label{fig:growth rate}
\end{figure}

One might expect accretion to have a more substantial influence in regions where the line mass is high. \citet{hacar2022initial}, for example, proposes that the normalisation of filaments in the M-L plot is set by accretion with high line masses seen in high accretion environments. Figure \ref{fig:growth rate} illustrates the relationship between the filament growth rate and the line mass and shows there is a potential relationship between the two in our models. We performed a linear regression on the logarithmic transformation of $m$ and $\Dot{m}$ to obtain the fit parameters. The standard error of the fit was calculated from the mean squared error, and the 99\% confidence intervals were derived by exponentiating the fitted values $\pm$ 2.58 times the standard error, providing an error band around our linear model. As discussed in Section \ref{sec:Accretion}, a positive filament growth rate can relate to both mass increasing and/or length shortening. To further investigate the role of accretion, we plotted a bar for each point, corresponding to $\dot{m}$ multiplied by a length correction factor $\frac{L_{m+1}}{L}$, to demonstrate the impact of length variation on $\dot{m}$. We observe that the corrected values tend to be closer to the fitted linear relationship. This suggests that accretion drives the line mass relation and that variations in length due to algorithmic variations introduce scatter on top.

\subsection{Projection effect of the three-dimension structure}

In the preceding sections, we have delved into the evolution pathways of filaments in the Mass-Length diagram and discussed the probable role of accretion. However, it is important to note that filament identification largely depends on two-dimensional projections of inherently three-dimensional structures. Consequently, this projection effect may introduce systematic errors in our understanding of filament properties and their evolution. In this section, we aim to discuss the implications of these projection effects on our analysis.

\begin{figure}
\includegraphics[width=0.53\textwidth]{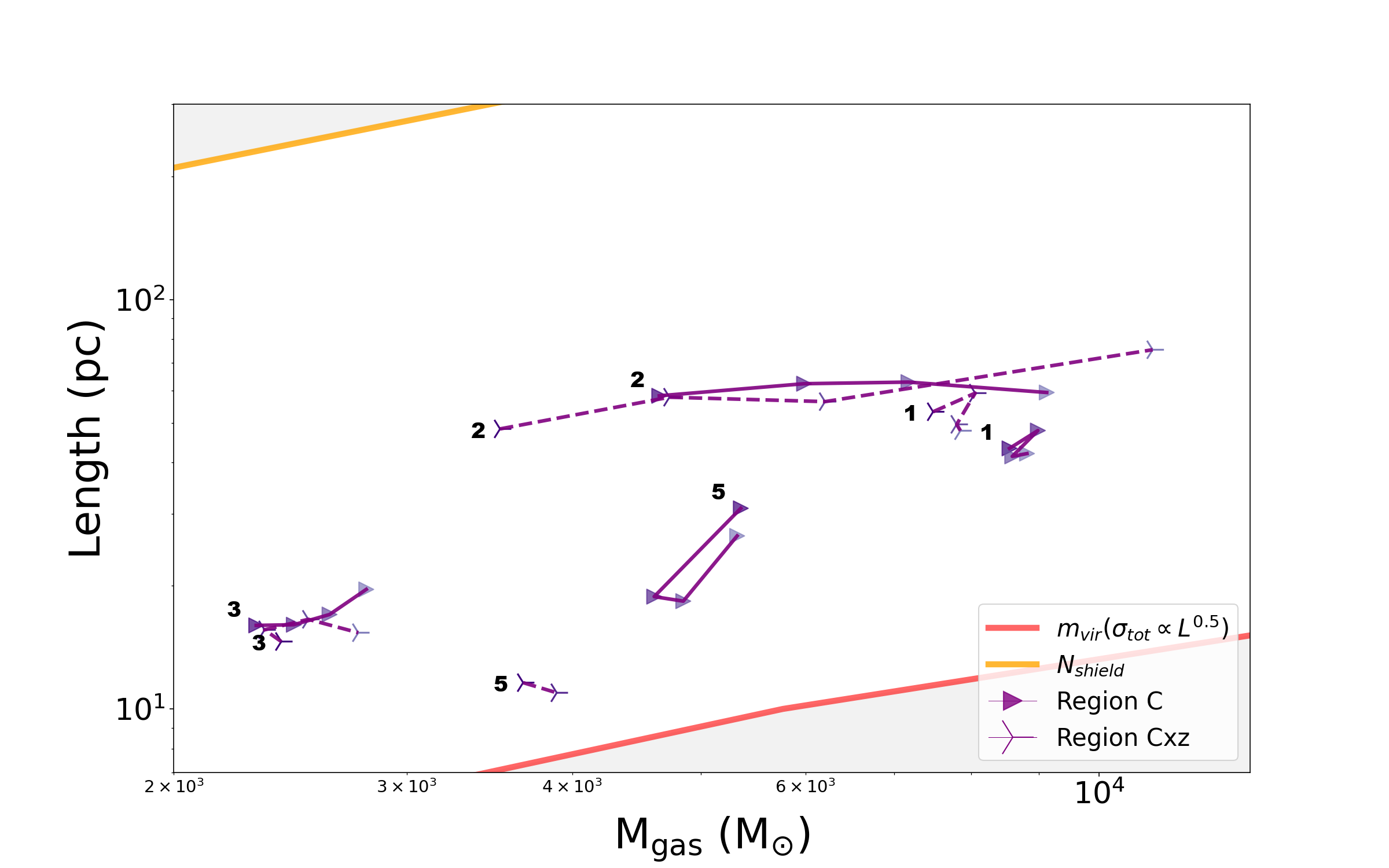}

\caption{The Mass-length correlations evolution of Region C filaments based on different view angle, same as Fig. \ref{fig:M-L all}.} 
\label{fig:M-L RC xz}
\end{figure}

In our previous radiative transfer realisations, the detector's line-of-sight is aligned along the Z-axis. As a result, the dust continuum images represent a projection onto the X-Y plane. We now simulate projections onto the X-Z plane by selecting the Y-axis as the new line-of-sight direction with \textsc{POLARIS} and applying the same analytical procedures. Figure \ref{fig:M-L RC xz} displays the evolutionary paths of filaments in different projections in Region C and shows that the length and mass of the filaments can vary due to projection. However, despite these variations, a similar trend in the evolutionary paths was observed for most filaments, except for RCF5. This consistency is largely due to the dispersed distribution of filaments in Region C, allowing the algorithm to identify individual filaments with ease.For RCF5, the example we use for the segmentation process in Section \ref{sec:M-L evolution}, the filament also became segmented in the X-Z plane.

However, RCF5 appears as multiple fragmented segments in the X-Z plane, making it difficult to correlate these parts with the more unified structure observed in the X-Y plane. Filaments in Region A are also difficult to correlate to individual structures across different projection planes. This is primarily because the molecular gas environment in Region A forms a continuous molecular complex. Areas of relatively low-density diffuse gas interconnect dense regions, significantly affecting how the algorithm identifies distinct filament structures in different projections. In other words, filaments of Region A identified in different projections are barely the same objects. These findings underscore the critical need to consider projection effects and the limitations of our current identification algorithm when analysing filamentary structures in diverse environments.

\subsection{Resolution effect: hierarchical filament structures?}
\label{sec:hierarchical}

The Mass-Radius relationship of molecular clouds has been extensively studied and is found to be linear, as evidenced by both observations and simulations \citep{larson1981turbulence, stutzki1998fractal, simon2001structure, roman2010physical, lombardi2010larson, qian201213co,kauffmann2010bmass}. The origin of this linear
relationship may be attributed to the hierarchical structure induced by turbulence and fragmentation \citep{kauffmann2010bmass}. Similarly, based on extensive observational data, \cite{hacar2022initial} have provided the mass-length relationship of filaments with approximately L $\propto$ M$^{0.5}$ and explored their hierarchical structure.

There are generally two mechanisms for the hierarchical structure of filaments. One mode of filament formation is the top-down process, where larger filaments fragment to form smaller filaments \citep{hacar2013cores}. The other mode is the bottom-up process, where small filaments are assembled onto pre-existing larger filaments under the influence of gravity to make them larger \citep{smith2014nature}. It is worth noting that both mechanisms may coexist. As we mentioned in Section \ref{sec:filament description}, the bimodal distribution of filament sizes is clearly seen in the PDF distribution in Figure \ref{fig:pdf} and can also be observed in the M-L diagram in Figure \ref{fig:M-L all}. The grey triangles in Figure \ref{fig:M-L all} can be clearly divided into two parts, one with a mass of up to $10^4 \ \rm M_{\odot}$ and the other with a size of around 10 pc and a mass ranging from $10^2$ to $10^3  \rm \  M_{\odot}$. However, due to the resolution limitation, we are not able to identify those smallest fibre-like structures with a size of around 0.1 pc. In most cases, especially in Regions B and C, the filaments are distinct and separate. This means we cannot link these filaments to the two mechanisms because we do not find any cases where filaments are within filaments.

\begin{figure}

\includegraphics[width=0.5\textwidth]{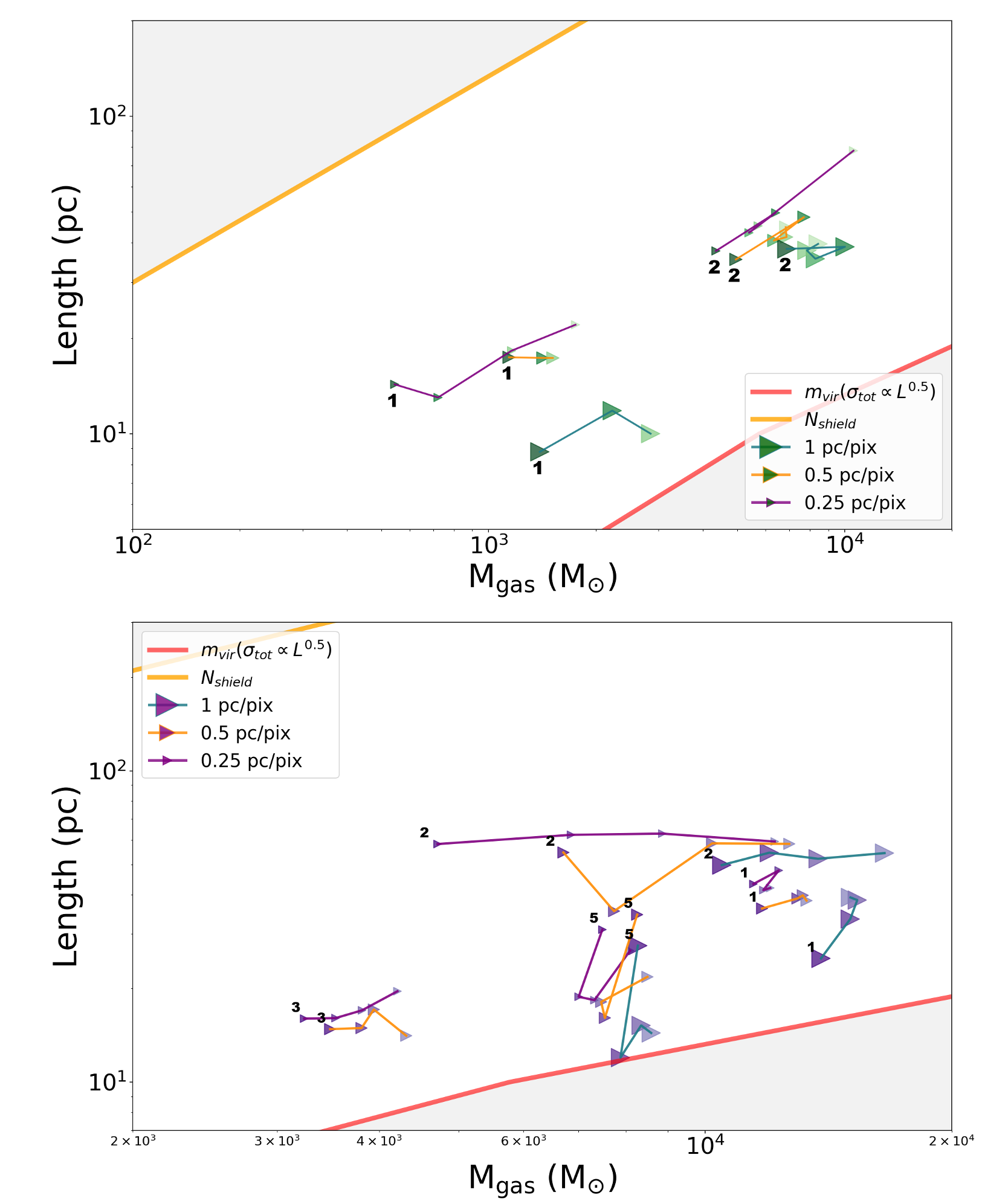}

\caption{The Mass-length correlations evolution of Region A (top) and Region C (bottom) filament in different resolutions, same as Fig \ref{fig:M-L all}. The marker sizes correspond to different resolutions: 0.25 pc/pix (small triangle), 0.5 pc/pix (medium triangle), and 1 pc/pix (large triangle). Filaments in different resolutions show a similar trend for the evolutionary path.} 
\label{fig:M-L diff res}
\end{figure}

Figure \ref{fig:M-L diff res} shows the tracked-filaments paths in different resolutions in Regions A and C. The filaments in Region C are clearly separated and distinct, providing a relatively simple environment for comparing the impact of resolution on their identification. Generally, the filaments identified at the lowest resolution have the highest masses, while those identified at the highest resolution have lower masses. The evolutionary paths show similarities across different resolutions, except for filament RCF1, where there is a large change in the mask.

\begin{figure}

\includegraphics[width=0.5\textwidth]{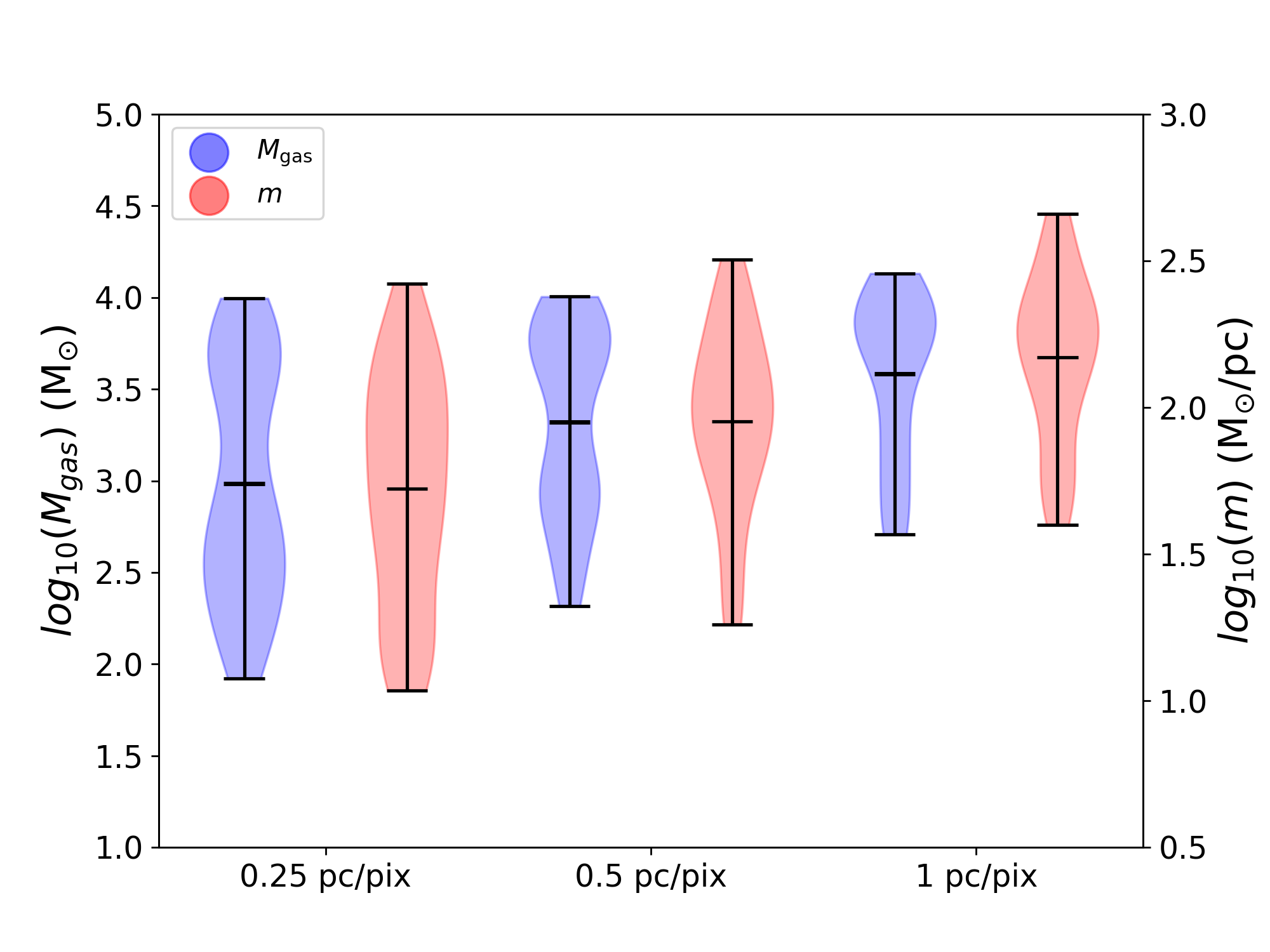}

\caption{The violin plot represent the distribution of filament properties $M_{\text{gas}}$ (blue) and $m$ (red) at different resolutions, with corresponding sample numbers of 82, 61, and 46 for 0.25, 0.5, and 1 pc/pix. The contours represent the density distribution, with central black lines indicating the mean values. The violin plot demonstrates the impact of spatial resolution on observed filamentary structures.}
\label{fig:resolution violin}
\end{figure}

Figure \ref{fig:resolution violin} illustrates the impact of resolution on the observed properties of filaments ($M_{\text{gas}}$ and $m$). As the resolution changes from 0.25 pc/pix to 1 pc/pix, there is a noticeable trend of increasing mean values for both $M_{\text{gas}}$ and $m$, with the predominant distribution of the violin plot contours shifting upward. This trend is consistent with the findings presented in Figure \ref{fig:M-L diff res}, corroborating the influence of spatial resolution on the characterization of filaments. In conclusion, the small structures are difficult for the algorithm to discern at low resolutions.
This explains how it is challenging to identify filaments with a mass distribution spanning several orders of magnitude in a single observation.
Still, when the statistics of all the filaments from different observations are combined and placed in the log-axes M-L diagram, these filaments exhibit a linear relationship. We conducted a linear regression on the log-transformed mass and length. In our case, we identified a fitted relation of $L \propto M^{0.45}$ as shown in Figure \ref{fig:M-L all}, which is indistinguishable from observations. The four mechanisms in Section \ref{sec:evo mechanism} maintain this linear relation with a dispersion that falls within the upper and lower limits predicted by the theoretical models.

\section{Conclusions}
We examined three molecular cloud regions in the Cloud Factory galactic-scale ISM suite, using synthetic dust observations using the radiative transfer code \textsc{POLARIS}. We observed the evolution of filaments, which are identified in the observational plane, with \textsc{FilFinder}.

We identified 189 filaments across the three regions at different resolutions in all snapshots and analysed their properties. We plotted their mass and length on an M-L plot and found a scaling relation $L \propto M^{0.45}$ similar to that seen in observations.
The distribution of our simulated filaments aligns well with that presented by \citet{hacar2022initial}. Comparisons with equilibrium models (virial line mass \(m_{\text{vir}}\) and critical line mass \(m_{\text{crit}}\)) suggest that most filaments are supercritical when considering only thermal support, making them susceptible to collapse and fragmentation. However, when turbulence is included, these filaments are found to be above the virial line, indicating a more complex stability balance.

We identified 9 filaments that could be tracked over an extended time period and studied their evolution process in a galactic-scale environment. Filaments are dynamic objects that can grow and fragment to form stars, as in the evolutionary paths shown in Figure \ref{fig:M-L all}. By studying their evolution, we found three main filament evolution mechanisms that dictate the evolutionary progression of filaments in the M-L phase diagram. Accretion leads to a mass increase, while length might increase or remain steady. Segmentation causes significant mass and length decreases, while dispersal leads to reduced length with little change in mass.

Most tracked filaments in our samples exhibit paths from smaller masses to larger masses in the M-L plot, indicating the dominant role of accretion in filament evolution. We also found a potential linear trend between filament growth rate and line mass, suggesting gravity-dominated accretion.

These three mechanisms determine the distribution of filaments in the M-L diagram. However, we acknowledge that internal dispersal mechanisms, such as stellar feedback from jets, outflows, and photoionization, are not included in our current simulation. Magnetic fields, which might help keep filaments more coherent and linear, are also not included. We plan to explore their impact in future research.

We investigate the effects of projection on filament evolution in the M-L diagram and find that it introduces systematic errors in the identification and analysis of filamentary structures. The general trend of the evolutionary paths of individual filaments in a simple environment is preserved in different projections. However, measurements like length and mass may vary. In more complex regions, projection affects the accuracy of our current identification algorithms, meaning the structures identified at different viewing angles do not correspond to each other. These findings emphasise the need for caution when interpreting results based solely on two-dimensional projections.

Finally, we inspect the hierarchical nature of the filaments. We compare the filaments identified at different resolutions. The higher-resolution images reveal more filaments with smaller masses, shorter lengths, and lower line masses. However, due to the resolution limit and algorithmic limitations, we could not find any instances of filaments nested within their parental filaments.

\section*{Acknowledgements}

We gratefully acknowledge helpful discussions with Andres Izquierdo, Jouni Kainulainen, Gina Panopoulou, Fabian Heitsch and Xuepeng Chen. J.F. acknowledges support of the National Natural Science Foundation of China (grant No. 12041305) and the CAS International Cooperation Program (grant No. 114332KYSB20190009), and grants from the STFC and CSC 201904910935, without which, this work would not have been possible. R.J.S. gratefully acknowledges an STFC Ernest Rutherford fellowship (grant ST/N00485X/1). A.H. acknowledges support and funding from the European Research Council (ERC) under the European Union’s Horizon 2020 research and innovation program (grant agreement No. 851435). S.E.C. acknowledges support from the National Science Foundation under grant No. AST-2106607. D.S. acknowledges support of the Bonn-Cologne Graduate School, which is funded through the German Excellence Initiative as well as funding by the Deutsche Forschungsgemeinschaft (DFG) via the Collaborative Research Center SFB 956 ``Conditions and Impact of Star Formation'' (subproject C6) and the SFB 1601 ``Habitats of massive stars across cosmic time” (subprojects B1 and B4). Furthermore, D.S. received funding from the programme “Profilbildung 2020", an initiative of the Ministry of Culture and Science of the State of Northrhine Westphalia.

This work used the DiRAC@Durham facility managed by the Institute for Computational Cosmology on behalf of the STFC DiRAC HPC Facility (www.dirac.ac.uk). The equipment was funded by BEIS capital funding via STFC capital grants ST/P002293/1, ST/R002371/1 and ST/S002502/1, Durham University and STFC operations grant ST/R000832/1. DiRAC is part of the National e-Infrastructure.

\section*{Data Availability}
Data supporting this study is available in the Zenodo repository at the following link: \href{https://zenodo.org/records/10117334}{https://zenodo.org/records/10117334}. The dataset encompasses the column density maps derived from the synthetic observations for all the snapshots across the three resolutions in three regions. Further data products are available upon request.


\bibliographystyle{mnras}
\bibliography{myref} 



\appendix

\section{Comparison between raw simulation and synthetic observation}
\label{sec:appendix}

When simulating synthetic observations with the \textsc{POLARIS} algorithm for radiative transfer, it's crucial to ensure the physical properties of the original data from the Cloud Factory is well-preserved. In this appendix, we conduct a thorough comparison between the original datasets and the data generated by \textsc{POLARIS}. The main goal of this comparison is to evaluate the reliability and relevance of the data after being simulated by \textsc{POLARIS}.

\begin{figure}
\includegraphics[width=0.5\textwidth]{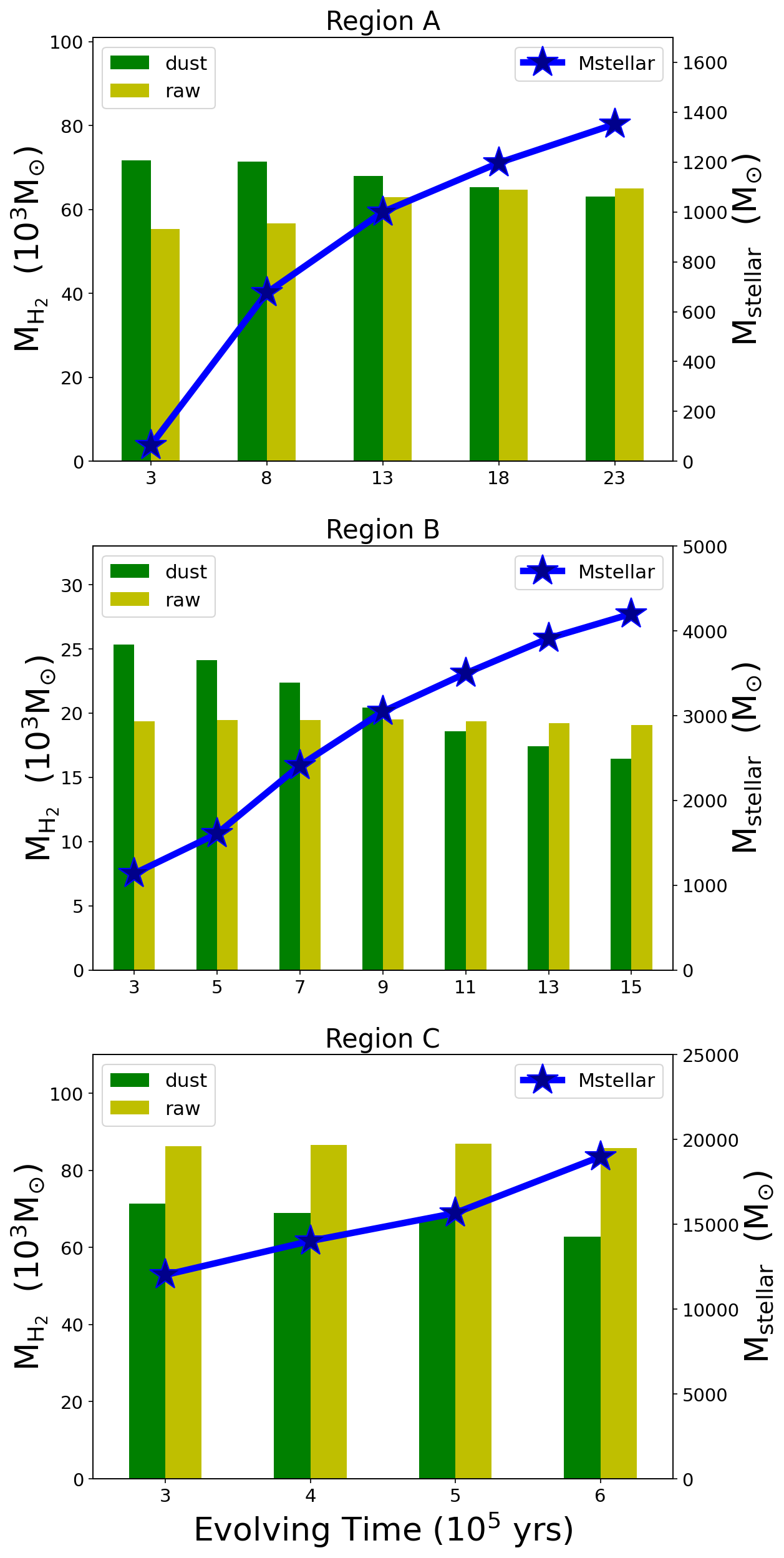}

\caption{The evolution of properties in three regions. The dark green bar show the synthetic observational mass $M\rm _{H_{2,dust}}$, and the shallow green bar show $M\rm _{H_{2,orig}}$ from the original simulation data. The blue lines show the evolved stellar mass $M_\mathrm{{stellar}}$. The stellar mass in all three regions increases over time, and also indicating a different star forming stage of three regions.} 
\label{fig:cloud properties}
\end{figure}

As shown in Figure \ref{fig:cloud properties}, $M\rm _{H_{2,dust}}$ and $M\rm _{H_{2,raw}}$ are not identical for the whole area. The stellar mass (33\% of the sink mass) increases in all regions with time as expected. However, the gas mass derived without radiative transfer $M_{\rm H_{2,raw}}$ shows little variation in regions B and C, and only a slight increase in region A. 
In the mean time, $M_{\rm H_{2,dust}}$ shows a decreasing trend. 
$M\rm _{H_{2,dust}}$ can be bigger or smaller than $M\rm _{H_{2,raw}}$, with a maximum difference of $\sim$ 30 \% seen in snapshot 3, of Region B. 

\begin{figure}
    \centering
    \includegraphics[width=1.05\linewidth]{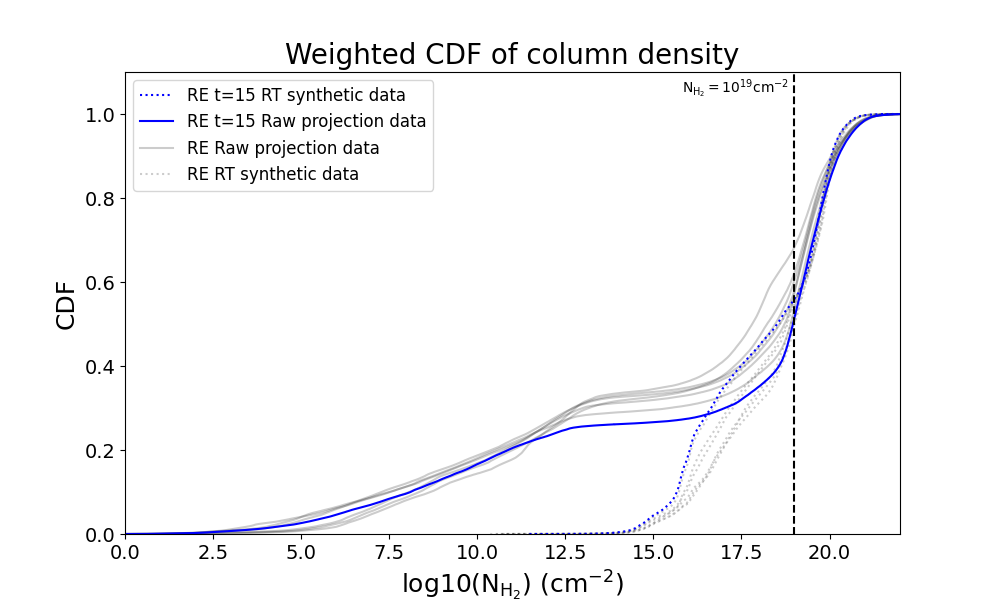}
    \caption{The figure illustrates the weighted cumulative distribution function (Weighted CDF) for column density $N_{\mathrm{H_2}}$ for data associated with Region B. The blue lines represent the data of snapshot 15 (time evolution of $15 \times 10^5$ yrs), with the solid line being the raw projection data from the Cloud Factory simulation, and the dotted line corresponding to the radiative transfer simulated observation data using the POLARIS algorithm. Other snapshots data are visualized using semi-transparent black lines. The black dashed vertical line denotes a reference column density of $10^{19} \rm \ cm^{-2}$. The two datasets converge and become strikingly similar for higher densities ($> 10^{19} \rm \ cm^{-2}$), indicating a good match between the RT synthetic data and the raw data in high-density regions.}

    \label{fig:cdf weighted}
\end{figure}

To further study the difference between the raw data and the radiative transfer simulation data, we generate projection column density maps from raw data and compare them to the dust-derived column density maps. Figure \ref{fig:cdf weighted} shows the weighted cumulative distribution function (Weighted CDF) for column density $N_{\mathrm{H_2}}$ for data associated with Region B (RB). It is evident from the figure that the weighted CDF exhibits a pronounced disparity between the raw and RT synthetic datasets for values below $10^{19}\ \rm cm^{-3}$. This distinction is manifested not only as a clear separation but also as an inverse trend observed over the course of evolving snapshots. However, in the relatively high column density which over $10^{19} \ \rm cm^{-3}$, the CDF of two datasets converge and show a similar trend, suggesting a good match in high-density regions. 

We also inspect the mass in the high-density regions, as shown in Table \ref{tab:high density mass}. It should be noted that it is difficult to compare the same area between the radiative transfer synthetic dust continuum data and the raw data, since the dust continuum data is two-dimension and the raw data is three-dimension. Moreover, the data format of the Cloud Factory is an irregular Voronoi grid, making it very difficult to match positions with dust data on a one-to-one basis. Therefore, we simply compared the $\rm H_2$ gas mass in high-density pixels/cells for column density ($10^{21}\ \rm cm^{-2}$) or volume density ($10^{2}\ \rm cm^{-3}$). Unlike comparing the total H$_2$ mass aggregated across all cells, we found certain similarities between the $M_{\rm H_{2},dust}$ and The $M_{\rm H_{2},raw}$ in the high-density area in regions A and C for the mass variation. Region B shows a different trend in early snapshots (3,5,7), however, we found that this difference disappears when we increase the volume density threshold to $10^3 \rm \ cm^{-3}$. This finding suggests the synthetic dust data fit well in the high-density region but distorted in the low-density region. In this work, since we only focus on the dense gas filament objects, which are the high-density regions, the distortion is acceptable. In our future work, we will further investigate the reasons for the inconsistencies between the radiative transfer simulation and the original data.

\begin{table}
\caption{Mass comparison between radiative transfer synthetic data and raw data. }
\begin{threeparttable}
{\small
\begin{tabular}{ccccc}
\hline
Region & Snapshot & $M_{\rm H_{2},dust}$ & $M_{\rm H_{2},raw}$  & $M_{\rm H_{2},raw} $  \\
\\
  &   & $(>10^{21} \rm \ cm^{-2})$ & $ (>10^2\rm \  cm^{-3})$ & All cells \\
 \hline
 &  & M$_{\odot}$  & M$_{\odot}$ & M$_{\odot}$  \\
        (1) & (2) & (3) & (4) & (5)  \\ \hline
        A & 3 & 22152 & 17822 & 55318 \\ 
        A & 8 & 25109 & 30967 & 56720 \\ 
        A & 13 & 21584 & 27896 & 62878  \\ 
        A & 18 & 16569 & 21577 & 64638  \\ 
        A & 23 & 13470 & 17282 & 64968 \\ 
        B & 3 & 3727 & 6595 & 19371 \\ 
        B & 5 & 3675 & 7198 & 19461  \\ 
        B & 7 & 3593 & 7579 & 19434  \\ 
        B & 9 & 3152 & 7550 & 19523  \\ 
        B & 11 & 2848 & 7183 & 19366 \\ 
        B & 13 & 2623 & 6451 & 19192 \\ 
        B & 15 & 2578 & 5700 & 19077  \\ 
        C & 3 & 31272 & 52963 & 86232  \\ 
        C & 4 & 31175 & 55022 & 86663  \\ 
        C & 5 & 29803 & 56105 & 86922  \\ 
        C & 6 & 27430 & 56444 & 85855 \\
 \hline
\end{tabular}
}
\begin{tablenotes}
\item[] \textbf{Note.} (1) Region of clouds (2) Evolving time snapshot, in units of 10$^5$ years (3) Total $\rm H_2$ mass in regions where the column density exceeds $10^{21} \ \rm cm^{-2}$, derived from RT synthetic data (4) Total H$_2$ mass accumulated from cells with a volume density greater than $10^{2} \ \rm cm^{-3}$, derived from raw data (5) Total H$_2$ mass aggregated across all cells, derived from raw data.
\end{tablenotes}
\label{tab:high density mass}
\end{threeparttable}
\end{table}

\bsp	
\label{lastpage}
\end{document}